\documentclass[times,sort&compress,3p]{elsarticle}
\usepackage[labelfont=bf]{caption}

\usepackage{amsmath,amsfonts,amssymb,amsthm,booktabs,color,epsfig,graphicx,hyperref,url}

\theoremstyle{plain}% Theorem-like structures provided by amsthm.sty

\newtheorem{proposition}{Proposition}

\theoremstyle{definition}

\usepackage{natbib}
\newtheorem{definition}{Definition}

\usepackage{amsmath}
\usepackage{graphicx}
\usepackage{enumerate}
\usepackage{natbib}
\usepackage{url} % not crucial - just used below for the URL 
\usepackage{latexsym}
\usepackage{amssymb}
\usepackage{booktabs} % To thicken table lines
\usepackage[export]{adjustbox} %%position of figure
\usepackage{hyperref}
\usepackage{multirow} %%%for the columns
\usepackage{float} %% pour les tables flottant
\usepackage{lscape}
\usepackage[english]{babel}
\usepackage{bm}
\def\spacingset#1{\renewcommand{\baselinestretch}%
{#1}\small} \spacingset{1.9}
\begin{document}
\def\spacingset#1{\renewcommand{\baselinestretch}%
{#1}\small} \spacingset{1.9}

\begin{frontmatter}

\title{\textbf{A new copula regression model for hierarchical data.}}

\author[1,2]{Talagbé Gabin Akpo}
\author[1]{Louis-Paul Rivest \corref{mycorrespondingauthor}}

\address[1]{Département de Mathematiques et de Statistique,  Université Laval, 2325, rue de l'Université, G1V 0A6, Québec, Canada.}

\address[2]{Institut National de la Recherche Scientifique, Centre Armand-Frappier Santé Biotechnologie, 531 boulevard des Prairies, H7V 1B7, Laval, Canada }

\cortext[mycorrespondingauthor]{Corresponding author. Email address : \url{Louis-paul.rivest@mat.ulaval.ca}}

\begin{abstract}
This paper proposes multivariate copula models for hierarchical data. They account for two types of correlation: one is between variables measured on the same unit and the other is a correlation between units in the same cluster. This model is used to carry out copula regression for hierarchical data that gives cluster specific prediction curves. In the simple case where a cluster contains two units and where two variables are measured on each one, the new model is constructed within a D-vine. Then we focus on situations where two variables are measured on the units of a cluster of arbitrary size. The proposed copula density has an explicit form; it is expressed in terms of three copula families. We study the properties of the model; compare it to the linear mixed model and end with special cases. When the three copula families and the marginal distributions are normal, the model is equivalent to a normal linear mixed model with random, cluster specific, intercepts. The method to select the three copula families and to estimate their parameters are proposed. We perform a Monte Carlo study of the parameter estimators.
A data set on the marks of students in several school is used to implement the proposed model and to compare its performance to standard normal mixed linear models.
\end{abstract}


\end{frontmatter}

\spacingset{1.9} % DON'T change the spacing!
\section{Introduction} \label{sec:intro}

A simple bivariate copula regression predicts dependent variable $y$ using independent variable $x$. It proceeds by selecting a bivariate copula summarizing the relationship between $x$ and $y$. The copula regression predictor is then constructed using characteristics of the conditional distribution of $y$ given $x$ derived from the selected copula, see \cite{Nel2006}, \cite{KumSho2007} and \cite{CraVan2008} exemplify this method while \cite{HohAnoBou2013} derive its asymptotic properties and \cite{AcaAzi2019} investigates predictions errors. This method is easily extended a multivariate explanatory variable $x$.

The goal of this work is to generalize the basic copula regression model to hierarchical data: $x$ and $y$ are observed on units that are in clusters and one would like to include a cluster effect in the copula regression predictions. The classical regression model for hierarchical data is a normal linear mixed model, with cluster specific random slopes and intercepts, see \cite{BatHatFul1988}, \cite{Ver2000}, \cite{McCulloch2001} and \cite{Gold2011}. The exchangeable copula families have been proposed to model the residual dependency within cluster in this context, see \cite{Riv2016} and \cite{grover2020copula}; \cite{SuNesWan2019} provide a survival data application of this approach.

A model for the joint distribution of all the $x$ and the $y$ variables in a cluster is first constructed. The conditions that the model must fulfill in order to yield suitable predictions are given in Section 2. It is required to meet an exchangeability assumption: permuting the units in a cluster does not change the joint distribution of the variables. It also relies on a partial conditional independence assumption that insures that the prediction of $y$ for a unit does not depend on the $x$ values for the other units in the cluster. The proposed model is then constructed within a D-vine in the simple case of a cluster containing two units. The general model, for clusters of arbitrary size, is introduced in Section \ref{section3}. Afterwards, we study its properties by showing that, in particular cases, conditional versions are equivalent to the models of \cite{BatHatFul1988}  and of \cite{Riv2016}. We use the proposed copula to do cluster specific predictions. The copula model is then implemented in a data set of \cite{Gold2011} and compared with standard normal mixed linear models.

\section{Model construction : Exchangeability and conditional independence}\label{section2}

This section considers that $d\ge 2$ variables are measured on all the units in a cluster; subscript $j$ represents a unit, $j=1,\ldots,n$, where $n$ is the size of the cluster. The dependent variable for unit $j$ is $Y_j$ while $\bm{X}_j$ is the corresponding vector of $d-1$ explanatory variables and $\bm{Z}_j=(\bm{X}_j^\top, Y_j)^\top$, is the vector of the $d$ variables measured on unit $j$. Let $F_{d,1:n}(\bm{z}_1,\ldots, \bm{z}_n)$ be the joint cumulative distribution function (cdf) of the $nd$ variables measured in the cluster, where $\bm{z}_j=(\bm{x}_j^\top,y_j)^\top$. The model is constructed using copulas; it is therefore of interest to define $F(\bm{x})=\left(F_1(x_1),\ldots,F_{d-1}(x_{d-1})\right)$ and $G(y)$ as the marginal distributions of respectively $\bm{X}_j$ and $Y_j$ which are the same for all the units. We let $\bm{U}_j=F(\bm{X}_j)$ and $V_j=G(Y_j)$ be random variables with uniform margins and $c_{d,1:n}\left\{(\bm{u}_1,v_1),\ldots,(\bm{u}_n,v_n)\right\}$ be the copula density for the joint distribution of $\{(\bm{U}_j^\top, V_j)^\top :\ j=1,\ldots,n\}$. We now give some conditions for the family of joint distributions $F_{d,1:n}(\bm{z}_1,\ldots, \bm{z}_{n}) :\ n=2,3,\ldots$ to give useful regression models.

\subsection{Exchangeability}\label{sec:exchang}
We consider the family of cumulative distribution functions defined by
\begin{equation*}
\mathcal{F}_{d}=\left\{F_{d,1:n}(\bm{z}_1,\ldots, \bm{z}_{n}) : \bm{z}_j \in \mathbb{R}^d, \,j=1,\ldots,n,\ n=1,2 \ldots \right\}\cdot
\end{equation*}

%\begin{definition}
The familly $\mathcal{F}_{d}$ is said to be $d$-exchangeable if, for all $n\ge 2$, $F_{d,1:n} \in \mathcal{F}_{d}$ satisfies the following conditions
\begin{itemize}
\item[$i)$] \textbf{Permutation invariance :} For all permutations $\{\pi(1), \ldots, \pi(n)\}$ of $\{1, 2, \ldots, n\}$,
\begin{equation} \label{eq:dexc1}
F_{d,1:n}\left(\bm{z}_{1},\ldots, \bm{z}_{n}\right)=
F_{d,1:n}\left\{\bm{z}_{\pi(1)},\ldots, \bm{z}_{\pi(n)}\right\}.
\end{equation}
\item[$ii)$] \textbf{Closure on marginalization:} For any $ r \leq n $,
\begin{equation}\label{eq:dexc2}
F_{d,1:r}\left(\bm{z}_{1},\ldots, \bm{z}_{r})=F_{d,1:n}(\bm{z}_{1},\ldots, \bm{z}_{r},\infty,\ldots,\infty\right).
\end{equation}
 \end{itemize}
%\end{definition}

This definition is similar to the classical definition of univariate exchangeability that is given in \cite{MaiSher2012}. A simple example of $d$-exchangeability is a multivariate one way ANOVA model with random effects, $\bm{Z}_j=\bm{A}+\bm{E}_j$, where $\bm{A}$ and $\bm{E}_j:\ j=1,\ldots, n$ are $d \times 1$ independent random vectors and $\{\bm{E}_j:\ j=1,\ldots, n\}$ have the same distribution. The next proposition gives the form of the correlation matrix for a $d$-exchangeable random vector.

\begin{proposition} \label{exch}
Let $\left\{\bm{Z}_1,\ldots,\bm{Z}_n\right\}$ be a set of $n d \times 1$ random vectors, verifying the definition of $d$-exchangeability given by \eqref{eq:dexc1} and \eqref{eq:dexc2}. Then, the $nd\times nd$ correlation matrix of Pearson, Spearman, and Kendall's between these $n$ vectors have the form
\begin{equation} \label{eq1}
I_{n} \otimes \Sigma_w + J_{n} \otimes \Sigma_b =
\left(
\begin{array}{cccc}
\Sigma_w + \Sigma_b & \Sigma_b & \dots & \Sigma_b\\
\Sigma_b & \Sigma_w + \Sigma_b & \ddots & \vdots\\
\vdots & \ddots & \ddots & \Sigma_b\\
\Sigma_b & \dots & \Sigma_b & \Sigma_w + \Sigma_b
\end{array}
\right),
\end{equation} 
where $J_{n}$ is and $n \times n$ matrix of ones and $\otimes$ denotes the Kroenecker product. Moreover, for Pearson and Spearman correlations, the matrices $\Sigma_w$ and $\Sigma_b$ are positive semi definite.
\end{proposition}

\subsection{Partial conditional independence}\label{sec:indep}\textbf{}\
This section proposes an assumption concerning the dependency within each cluster.
\begin{definition}
The random variables $Y_j$ and $\{\bm{X}_k:\ k\ne j\}$  are assumed to be independent, given $\bm{X}_j$. This is a partial conditional independence assumption that can be written as
\begin{equation}\label{indepenpar}
Y_j \perp \{\bm{X}_k:\ k\ne j\} |\bm{X}_j, \ j=1,\ldots,n.
\end{equation}
This condition is weaker than the conditional independence assumption underlying the standard regression model that can be formulated as 
\begin{equation*}\label{indepenc}
Y_j \perp \{(\bm{X}_k,Y_k):\ k\ne j\} |\bm{X}_j, \ j=1,\ldots,n.
\end{equation*}
\end{definition}

We now implement the exchangeability and the independence assumption within a $D$-vine for the joint distribution of two units within a cluster when $d=2$, that is when $\bm{Z}=(X,Y)^\top$.

\subsection{A D-vine construction of the model when $d=n=2$}\label{sec:indep1}
\begin{figure}[H]
 \centering
 \includegraphics[scale=0.7,height=8.0cm,width=13cm]{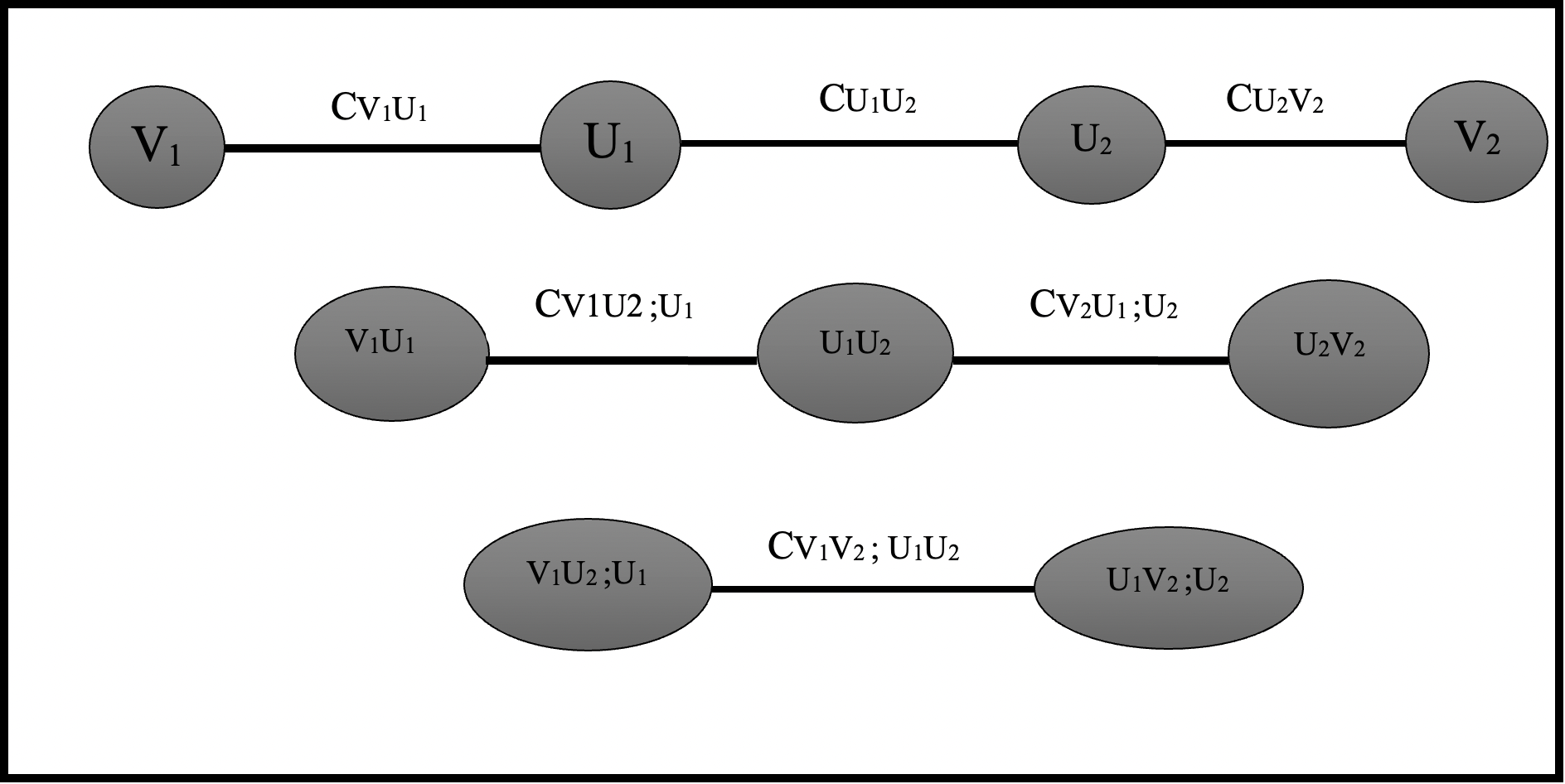}
 \caption{Graphical representation of the proposed $D$-vine.}
 \label{fig:my_label1}
\end{figure}

\textbf{}\
This section constructs a copula density function for the random variables in a cluster containing two units and verifying the properties of partial conditional independence and 2-exchangeability. The 4 random variables are $U_1=F(X_1)$, $V_1=G(Y_1)$, $U_2=F(X_2)$ and $V_2=G(Y_2)$. The four trees for the proposed $D$-vine are given in Figure \ref{fig:my_label1}. 

The decomposition involves 6 bivariate copulas $C_{U_1V_1}$, $C_{U_1U_2}$, $C_{U_2V_2}$, $C_{V_1U_2; U_1}$, $C_{U_1V_2 ; U_2}$ and $C_{V_1V_2 ; U_1 U_2}$. The density of the multivariate copula corresponding to the trees given in Figure \ref{fig:my_label1} can be derived from \citet[pp. 108]{Joe2014a}. It is given by
\begin{eqnarray}\label{eq0}
c\left\{(u_1,v_1),(u_2,v_2)\right\}&=&c_{U_1V_1}(u_1,v_1)c_{U_2V_2}(u_2,v_2)c_{U_1U_2}(u_1,u_2) c_{V_1U_2; U_1}\left\{C_{V_1|U_1}(v_1|u_1), C_{U_2|U_1}(u_2|u_1)\right\}\nonumber\\
&\times&  c_{U_1V_2 ; U_2}\left\{C_{U_1|U_2}(u_1|u_2), C_{V_2|U_2}(v_2|u_2)\right\} c_{V_1V_2; U_1 U_2}\left\{C_{V_1|U_1U_2}(v_1|u_1,u_2), C_{V_2|U_1U_2}(v_2|u_1,u_2)\right\},
\end{eqnarray}
where the conditional distribution functions $C_{V|U}$ and $C_{W|UV}$ are defined by
\begin{equation}\label{dcondi1}
C_{V|U}(v|u)=\frac{C_{UV}(u,v)}{\partial u},\,\,\,C_{W|UV}(w|u,v)=\frac{\partial C_{WV;U}\left\{C_{V|U}(v|u),C_{W|U}(w|u)\right\}}{\partial C_{V|U}(v|u)}\cdot
\end{equation}
This vine is assumed to fulfill the simplifying assumption: the copulas associated to the second and third trees do not depend on the conditioning variables. Discussions of this vine decomposition can be found in \cite{AnaCzaJoe2012}, \cite{Dis2013}, and \cite{CzaNag2022}. We would like the density \eqref{eq0} to fulfill the conditions for 2-exchangeability, see \eqref{eq:dexc1} and \eqref{eq:dexc2}, and for partial conditional independence, see \eqref{indepenpar}.\\
The density of the copula is 2-exchangeable if $c\left\{(u_1,v_1),(u_2,v_2)\right\}=c\left\{(u_2,v_2),(u_1,v_1)\right\}$. This requires a unique copula density for the dependency between $U$ and $V$, that is $c_{U_1V_1}(u,v)=c_{U_2V_2}(u,v)$, for $u,v \in (0,1)$. Also, the copula for the dependency between $U_1$ and $U_2$ needs to be symmetric, that is $c_{U_1U_2}(u,v)=c_{U_1U_2}(v,u)$ and $c_{V_1V_2 ; U_1U_2}(u,v)=c_{V_1V_2 ; U_1U_2}(v,u)$. This definition also entails restrictions on the conditional copula densities $c_{V_1U_2|U_1}$ and $c_{U_1V_2| U_2}$. However for (\ref{eq0}) to fulfill the partial conditional independence condition \eqref{indepenpar} these two copulas need to be equal to the independence copula. This leads to following copula density
\begin{eqnarray} \label{eq:deq2}
c\{(u_1,v_1),(u_2,v_2)\}&=& c_{U_1U_2}(u_1,u_2) \times \left\{c_{U_1V_1}(u_1,v_1) c_{U_1V_1}(u_2,v_2)\right\}c_{V_1V_2 ;U_1U_2}\{C_{V_1|U_1}(v_1|u_1),
C_{V_1|U_1}(v_2|u_2)\}\cdot
\end{eqnarray}
It involves an arbitrary copula density $c_{UV}$ for the relationship between $U$ and $V$ and two symmetric copula densities, $c_{U_1U_2}$ and $c_{V_1V_2 ; U_1U_2}$.

It is interesting to consider the special case where the three copulas in \eqref{eq:deq2} are normal. Let $\rho_1$ represent the correlation between $U_1$ and $U_2$, $\rho_2$ that between $U$ and $V$ and $\rho_3$ be the residual correlation. The joint copula density \eqref{eq:deq2} for $(U_1,U_2,V_1,V_2)$ is normal, see \citet[pp. 119]{Joe2014a}. Its correlation matrix is given by

\begin{equation} \label{coro1}
\left(
\begin{array}{cccc}
1 & \rho_1 & \rho_2 & \rho_1\rho_2\\
\rho_1 & 1 & \rho_1\rho_2 & \rho_2\\
\rho_2 & \rho_1\rho_2 & 1 & \rho_1\rho_2^2+\rho_3(1-\rho_2^2)\\
\rho_1\rho_2 & \rho_2 &\rho_1\rho_2^2+\rho_3(1-\rho_2^2) & 1
\end{array}
\right)\cdot
\end{equation}
In \eqref{coro1}, the result that $\mathbb{E}\left\{\Phi^{-1}(U_1)\Phi^{-1}(V_2)\right\}=\rho_1\rho_2$ is obtained by conditioning on $U_2$ while the variance-covariance matrix of $\left\{\Phi^{-1}(V_1),\Phi^{-1}(V_2)\right\}$ knowing $(U_1,U_2)$ is used to find that $\rho_3=\left\{\mathbb{E}\left\{\Phi^{-1}(V_1)\Phi^{-1}(V_2)\right\}-\rho_1\rho_2^2\right\}/(1-\rho_2^2)$.

Given that $(U_1,U_2)$ are distributed according to a normal copula with correlation $\rho_1$, one obtains the normal copula with correlation matrix \eqref{coro1} for the joint distribution of $(U_1,U_2,V_1,V_2)$ if $V_1$ and $V_2$ are defined by
\begin{equation}\label{decompbat}
\Phi^{-1}(V_j)=\rho_2\Phi^{-1}(U_j)+\sqrt{\rho}_3(1-\rho_2^2)^{1/2}A+(1-\rho_2^2)^{1/2}(1-\rho_3)^{1/2}E_j,\,\,j=1,2,
\end{equation}
where $A, E_1, E_2$ are independent with a $N(0,1)$ distribution. Model \eqref{decompbat} is similar to the linear mixed model of \cite{BatHatFul1988}. The term $\rho_2\Phi^{-1}(U_j)$ represents the contribution of the explanatory variables to the regression; it is fixed if one conditions on $(U_1,U_2)$. The cluster specific random intercept, $\sqrt{\rho}_3(1-\rho_2^2)^{1/2}A$, is independent of the experimental error, $(1-\rho_2^2)^{1/2}(1-\rho_3)^{1/2}E_j$. One can easily show that the correlation matrix of $\{\Phi^{-1}(U_1),\Phi^{-1}(U_2),\Phi^{-1}(V_1),\Phi^{-1}(V_2)\} $ entering in \eqref{decompbat} is given by \eqref{coro1}. Note also that \eqref{decompbat} is easily generalized to clusters of size $n>2$. This is also true of the general model \eqref{eq:deq2}. This leads to the general 2-exchangeable model that is proposed in the next section.

\section{A multivariate 2-exchangeable copula model}\label{section3}\textbf{}\
This section proposes a method to construct densities for the joint distribution of $\bm{Z}_j=({X}_j, Y_j)^\top$, for $j=1,\dots,n$ that meets constraints of 2-exchangeability and of partial conditional independence presented in Section \ref{section2}. The joint cdf for $\{\bm{Z}_j:\ j=1,\ldots,n\}$ is denoted $F_{2,1:n}$, where the first index, 2, refers to the dimension of $\mathbf{Z}$ while the index $1:n$ means that it concerns $n$ units labelled from 1 to $n$.
This joint distribution involves marginal distributions $F(x)$ and $G(y)$ for $X$ and $Y$ and a copula density 
$c_{2,1:n}\left\{(u_1,v_1),\ldots, (u_n,v_n)\right\}$. This density depends on $c^{(2)}$, a copula density for the relationship between $X$ and $Y$, and two families of 1-exchangeable copula densities, $\{c_{1,1:n}^{(1)}(u_1,\ldots,u_n):\ n \ge 2\}$ and $\{c_{1,1:n}^{(3)}(u_1,\ldots,u_n):\ n \ge 2\}$ for the dependency between the $X$ variables and the copula regression residuals within a cluster respectively. These two families are assumed to fulfill conditions \eqref{eq:dexc1} and \eqref{eq:dexc2} for $d=1$. 

The general form for the 2-exchangeable copula density is
\begin{eqnarray} \label{eq:ec1}
c_{2,1:n}\left\{(u_1,v_1),\ldots, (u_n,v_n)\right\}&=&c_{1,1:n}^{(1)}(u_1,\ldots,u_n)\times\overset{n}{\underset{j=1}{\prod}}\left\{c_{}^{(2)}(u_j,v_j)\right\}\times c_{1,1:n}^{(3)}\left\{C_{2|1}(v_1|u_1),\ldots,C_{2|1}(v_n|u_n)\right\},
\end{eqnarray}
where $u_j, v_j\in [0,1]$, $j=1,\ldots,n$ and the conditional distribution $C_{2|1}$ is deduced from equation \eqref{dcondi1} with $C_{UV}$ replaced by $C_{}^{(2)}$.
For $n=2$, equation \eqref{eq:ec1} reduces to the $D$-vine copula density \eqref{eq:deq2}
%; it fulfills the conditions of partial conditional independence and of 2-exchangeability.

% \begin{Prop}
% The density $c_{2,1:n}$ is a positive, closed function on the margins satisfying the multidimensional exchangeability property in dimension $d$ and the condition of partial conditional independence of the equation \eqref{indepenpar}.
% \end{Prop}

% \begin{proof}
To prove that \eqref{eq:ec1} meets the conditions \eqref{eq:dexc1} and \eqref{eq:dexc2} for 2-exchangeability and \eqref{indepenpar} for conditional independence we integrate the proposed joint density for $\{(U_1,V_1),\ldots, (U_n,V_n)\}$ in \eqref{eq:ec1} for $v_n\in (0,1)$. To carry this out, it is convenient to change variable, $w_n=C_{2|1}(v_n|u_n)$. The jacobian is $dw_n=c_{}^{(2)}(u_u,v_n)dv_n$. Using the closure on marginalization property of copula family $C_{1,1:n}^{(3)}$, the integral is equal to
\begin{eqnarray}\label{eq:inter}
c_{1,1:n}^{(1)}(u_1,\ldots,u_n)\times\overset{n-1}{\underset{j=1}{\prod}}c_{}^{(2)}(u_j,v_j)\times c_{1,1:(n-1)}^{(3)}\left\{C_{2|1}(v_1|u_1),\ldots,C_{2|1}(v_{n-1}|u_{n-1})\right\}\cdot
\end{eqnarray}
The variable $u_n$ only appears in $c_{1,1:n}^{(1)}(u_1,\ldots,u_n)$. Using the closure on marginalization property of copula family $C_{1,1:n}^{(1)}$, the integral on $u_n$ gives the density \eqref{eq:ec1} for the $(n-1)$ pairs $(U_1,V_1), \ldots, (U_{n-1},V_{n-1})$. Thus \eqref{eq:ec1} defines a proper copula density that meets requirements \eqref{eq:dexc1} and \eqref{eq:dexc2} for 2-exchangeability. To prove the partial conditional independence assumption \eqref{indepenpar} one integrates \eqref{eq:inter} for $v_{n-1}, v_{n-2},\ldots, v_2 \in (0,1)$. This is easily carried by changing variables,
$w_j=C_{2|1}(v_j|u_j), \ j=2,\ldots, n-1$. The joint density of $(V_1, U_1,\ldots,U_n)$ is given by 
$c_{1,1:n}^{(1)}(u_1,\ldots,u_n)c_{}^{(2)}(u_1,v_1)$; thus, given $U_1$, $V_1$ and $(U_2,\ldots, U_n)$ are independent and \eqref{indepenpar} holds.
% \end{proof}

The derivations in the previous paragraph have highlighted a key property of the proposed model. If the density of\\ $\left\{(U_1,V_1), \ldots , (U_n,V_n)\right\}$ is \eqref{eq:ec1} then the two vectors
$(U_1,\ldots,U_n)$ and $\{W_1=C_{2|1}(V_1|U_1), \ldots,W_n=C_{2|1}(V_n|U_n)\} $ are independent with densities respectively given by $c_{1,1:n}^{(1)}(u_1,\ldots,u_n)$ and $c_{1,1:n}^{(3)}(w_1,\ldots,w_n)$. This is summarized in the following proposition.

\begin{proposition}\label{theoind}
Let $(U_1,V_1), \ldots, (U_n,V_n)$, be a set of $n$ random vectors, whose joint density is given by \eqref{eq:ec1}. If $W_j=C_{2|1}(V_j|U_j),\,\,j=1,\ldots,n$, where the distribution function $C_{2|1}$ comes from \eqref{dcondi1}, then the random vectors $(U_1,\ldots,U_n)$ and $(W_1,\ldots,W_n)$ are independent with respective densities $c_{1,1:n}^{(1)}$ and $c_{1,1:n}^{(3)}$.
\end{proposition}

%\begin{proof} 
%Let $(U_1,V_1), \ldots, (U_n,V_n)$ be a set of $n$ random vectors where each is of %dimension $2$ and has a copula density that satisfies the definition of the $2$-dimensional exchangeable copula in dimension $n$. Using the formula for multivariate change of variables, the determinant of the Jacobian matrix of the random vector $(U_1,\ldots,U_n,W_1,\ldots,W_n)$ where $W_i=C_{2|1}(V_i|U_i)$ is
%\begin{equation*}
%|J|=c_{}^{(2)}(u_1,v_1)\times \ldots \times c_{}^{(2)}(u_n,v_n)\cdot
%\end{equation*}
%The density of the random vector is $c(u_1, \ldots, u_n, w_1,\ldots,w_n)=c_{1,1:n}^{(1)}(u_1, \ldots, u_n) \times c_{1,1:n}^{(3)}\left(w_1,\ldots,w_n\right)$. This concludes the proof of the theorem on the independence between the $n$-dimensional random vector $(U_1,\ldots, U_n)$ and the $n$-dimensional random vector $(W_1,\ldots,W_n)$ and gives the distribution of each of the random vectors.
%\end{proof}

The result of Proposition \ref{theoind} suggests the following algorithm to simulate a random vector with a density given of \eqref{eq:ec1}:

\begin{itemize}
\item Step 1 : Simulate $(U_1,\ldots,U_n)$ according to the exchangeable copula $C_{1,1:n}^{(1)}(u_1,\ldots,u_n)$;
\item Step 2 : Simulate $(W_1,\ldots,W_n)$ according to the exchangeable copula  $C_{1,1:n}^{(3)}(w_1,\ldots,w_n)$; 
\item Step 3 : Solve, the equations in $V_j$ defined by
 $W_j=C_{2|1}(V_j|U_j),\,\,\,\,\,j=1,\ldots,n$, where $C_{2|1}$ comes from the equation \eqref{dcondi1}. The the joint density of $(U_1,V_1,\ldots, U_n,V_n)$ is given by \eqref{eq:ec1}.
\end{itemize}
This algorithm differs from the proposal of \citet[pp. 136]{Caz2019} to simulate from a $D$-vine, that goes through the vine sequentially.

It is interesting to construct the joint density of $\{(X_1,Y_1),\ldots,(X_n,Y_n)\}$ from \eqref{eq:ec1}. It involves the joint marginal density of $(X_1,\ldots,X_n)$, 
$$ g_{1,1:n}^{(1)}(x_1,\ldots,x_n)=c_{1,1:n}^{(1)}\left\{F(x_1),\ldots,F(x_n)\right\}\prod_{j=1}^nf(x_j) \cdot$$
The conditional densities of $Y_j$ given $X_j$, for $j=1,\ldots,n$,
$$\prod_{j=1}^n \left[g(y_j)c_{}^{(2)}\left\{G(y_j),F(x_j)\right \}\right],$$
and a term for the residual dependency within clusters:
$$c_{1,1:n}^{(3)}\left[C_{2|1}\left\{G(y_1)|F(x_1)\right\},\ldots,C_{2|1}\left\{G(y_n)|F(x_n)\right\}\right] \cdot$$
This highlights a step wise construction of the 2-exchangeable model: first comes the specification of the marginal distribution for $X$, then that for the conditional distribution of $Y$ given $X$ and one finally adds a component for the residual dependency within a cluster.
% \begin{corollary}
% If the copula $C_{1,1:n}^{(3)}$ is an independence copula, we obtain the model proposed by \cite{Riv2016}. Moreover, if the marginal laws are normal distributions, the copulas $C^{(2)}$ and $C_{1,1:n}^{(3)}$ are normal then the 2-changeable copula model is equivalent to the linear mixed model, see \cite{BatHatFul1988}.
% \end{corollary}
%\begin{proof}
% We assume that the copula $C_{1,1:n}^{(3)}$ is an independence copula. By replacing in the equation \eqref{eq:ec1}, we obtain the model specified by \cite{Riv2016}.\\

We now assume that the copula $C^{(2)}$ is normal with correlation $\rho_2$. The conditional distribution in \eqref{eq:ec1} and its inverse are given by
$$ 
C_{2|1}(v|u)=\Phi\left\{\frac{\Phi^{-1}(v)-\rho_2\Phi^{-1}(u)}{\sqrt{1-\rho_2^2}}\right\}
\mbox{~~and~~}C_{2|1}^{-1}(t|u)=\Phi\left\{\Phi^{-1}(t)\sqrt{1-\rho_2^2}+\rho_2\Phi^{-1}(u)\right\},
$$
for $u, v, t \in [0,1]$, see \cite{BerCsa2015} for similar results. Using Proposition \ref{theoind}, the dependent variable $Y_j$ for a unit in a cluster of size $n$ with explanatory variable $x_j$ can be expressed as $Y_j=G^{-1}[C_{2|1}^{-1}\{W_j|F(x_j)\}]$, where $W_j$ is the entry of a vector with distribution $C_{1,1:n}^{(3)}$. This gives the following linear model :
$$
\Phi^{-1}\{G(Y_j)\}= \rho_2\Phi^{-1}\{F(x_j)\} + \sqrt{1-\rho_2^2}\Phi^{-1}(W_j)\cdot
$$
If, in addition, the marginal distributions of $X$ and $Y$ are normal: $F(y)=\Phi\{(x-\mu_1)/\sigma_1\}$ and $G(y)=\Phi\{(y-\mu_2)/\sigma_2\}$ where $\mu_1,\mu_2, \sigma_1^2, \sigma_2^2$ are respectively the marginal means and variances, then the model becomes:
\begin{equation} \label{eq:condmod}
Y_j= \beta_0+ \beta_1 x_j + \sigma_e\Phi^{-1}(W_j),
\end{equation}
where $\beta_0=\mu_2+\beta_1\mu_1$, $\beta_1=\rho_2\sigma_2/\sigma_1$, and $\sigma_e^2=\sigma_2^2(1-\rho_2^2)$. In \eqref{eq:condmod} the conditional marginal distribution of $Y_j$ is normal. 
%\end{proof}

The joint distribution depends on the copula $C_{1,1:n}^{(3)}$. These models are investigated in \cite{Riv2016}. If this copula is normal and exchangeable, with correlation $\rho_3$, then \eqref{eq:condmod} reduces to the normal mixed model of \cite{BatHatFul1988}. Finally note that the conditional density in \eqref{eq:condmod} for $(Y_1,\ldots,Y_n)$ given $(X_1=x_1,\ldots,X_n=x_n)$ is given by
\begin{eqnarray*} 
f(y_1,\ldots,y_n|x_1,\ldots, x_n)= \frac 1 {(2\pi)^{n/2} \sigma_e^n}\exp\left\{-\frac 1{2\sigma_e^2}\sum_{j=1}^n(y_j-\beta_0-\beta_1 x_j)^2\right\}
 c_{1,1:n}^{(3)}\left[\Phi\left\{(y_1-\beta_0-\beta_1 x_1)/\sigma_e\right\},\ldots,\Phi\left\{(y_n-\beta_0-\beta_1 x_n)/\sigma_e\right\}\right]\cdot\nonumber
\end{eqnarray*}
Thus model \eqref{eq:condmod} can easily be fitted by maximum likelihood.

\subsection{Predictions with the 2-exchangeable copula model}\label{section41}\textbf{}\
Suppose that $(n-1)$ units $\left\{(x_1,y_1),\ldots, (x_{n-1},y_{n-1})\right\}$ have been observed in a cluster. This section investigates the conditional distribution of $Y_{n}$ given $x_{n}$ in that cluster. A closed form expression for the conditional expectation of $Y_{n}$ given $x_{n}$ and $\left\{(x_1,y_1),\ldots, (x_{n-1},y_{n-1})\right\}$ is derived. Illustrations of the prediction curves for various specifications of the copula $C^{(2)}$ are presented.

The conditional density, $g_P(y_n)$ of $Y_{n}$ given $x_{n}$ and $\left\{(x_1,y_1),\ldots, (x_{n-1},y_{n-1})\right\}$ is expressed in terms of $w_j=C_{2|1}\left\{G(y_{j})|F(x_{j})\right\}$, $j=1,\ldots,n-1$. It is given by the ratio of \eqref{eq:ec1} over \eqref{eq:inter}, this yields
\begin{eqnarray}\label{eq:pred1}
g_P(y_n)=g(y_{n})c_{}^{(2)}\left\{F(x_{n}),G(y_{n})\right\}
\frac{c_{1,1:n}^{(3)}\left[w_1,\ldots,w_{n-1},C_{2|1}\left\{G(y_{n})|F(x_{n})\right\}\right]}{c_{1,1:(n-1)}^{(3)}(w_1,\ldots,w_{n-1})}\cdot
\end{eqnarray}
Observe that the conditional density of $W=C_{2|1}\left\{G(Y_{n})|F(x_{n})\right\}$, given
$\left\{(x_1,y_1),\ldots, (x_{n-1},y_{n-1})\right\}$, is simply\\
$c_{1,1:n}^{(3)}(w_1,\ldots,w_{n-1},w)/c_{1,1:(n-1)}^{(3)}(w_1,\ldots,w_{n-1})$. This is the conditional distribution of $(W_n|W_1,\ldots,W_{n-1})$, when $(W_1,\ldots,W_n)$ is distributed according to copula $C_{1,1:n}^{(3)}$. Thus one can easily simulate from \eqref{eq:pred1}. Starting from the equation \eqref{eq:pred1} and making a change of variable $w=C_{2|1}\left\{G(y_n)|F(x_n)\right\}$, we easily obtain the result.

 %\begin{proposition}
The best predictor of the unknown $Y_n$ is its conditional expectation, given $\left\{x_{n},(x_1,y_1), \ldots, (x_{n-1},y_{n-1})\right\}$, where\\ $w_j=\left[C_{2|1}\left\{G(y_j)|F(x_j)\right\} \right]$ it can be expressed as 
\begin{eqnarray}
\mathbb{E}_P(Y_{n})&=&
\int_0^1G^{-1}\left[C_{2|1}^{-1}\left\{w|F(x_{n})\right\}\right]\frac{c_{1,1:n}^{(3)}(w_1,\ldots,w_{n-1},w)}{c_{1,1:(n-1)}^{(3)}(w_1,\ldots,w_{n-1})}dw \cdot\label{eq:pred11}
\end{eqnarray}
 %\end{proposition}
%\begin{proof}
%\end{proof}
If the copula $C_{1,1:n}^{(3)}$ is the independence copula this reduces to a standard, unconditional copula regression for $C^{(2)}$, see \cite{KumSho2007}, \cite{CraVan2008} or \cite{HohAnoBou2013}. Taking the expectation of $\mathbb{E}_P(Y_{n})$ with respect to the distribution of $\left\{(X_1,Y_1),\ldots, (X_{n-1},Y_{n-1})\right\}$ also gives the unconditional copula regression curve for $C^{(2)}$. Thus the proposed model gives regression curves that vary between clusters and their expectation is equal to the marginal copula regression based on $C^{(2)}$.

%\begin{corollary}
%\begin{proof}
We suppose that the copula $C_{1,1:n}^{(3)}$ is exchangeable normal with correlation $\rho_3$. 
Thus if $W_j=C_{2|1}\left\{G(Y_j)|F(x_j)\right\},\,\,j=1,\ldots,n$, then the cumulative distribution function of random vector $\left\{\Phi^{-1}(W_1),\ldots,\Phi^{-1}(W_n)\right\}$ is a multivariate normal distribution with an exchangeable correlation matrix $\Sigma(n,\rho_3)$ whose entries are 1 on the diagonal and $\rho_3$ off the diagonal. Using standard properties of the multivariate normal distribution, the conditional distribution of $\Phi^{-1}(W_n)$ knowing $\left\{\Phi^{-1}(W_1),\ldots,\Phi^{-1}(W_{n-1})\right\}$ is univariate normal with mean $\mu_0$ and variance $\sigma^2_0$ defined by 
\begin{equation} \label{condimom}
\mu_0=\frac{(n-1)\rho_3 \sum_{j=1}^{n-1}\Phi^{-1}(W_j)/(n-1)}{1+(n-2)\rho_3},\,\,\,\,\,\,\,\,
\sigma_0^2=\frac{(1-\rho_3)\left\{1+(n-1)\rho_3\right\}}{1+(n-2)\rho_3},
\end{equation}
see for instance, \cite{Riv2016}.
%\end{corollary}
Thus the conditional distribution of random variable $\Phi^{-1}(W_n)=\Phi^{-1}[C_{2|1}\{G(Y_n)|F(x_n)\}$ is a $N(\mu_0,\sigma_0^2)$. The final form of \eqref{eq:pred11} when $C_{1,1:n}^{(3)}$ is a normal copula is therefore given by
\begin{equation}\label{finalpred}
 \mathbb{E}_{P}(Y_{n}) =\int_\mathbb{R} G^{-1}\left[C_{2|1}^{-1}\left\{\Phi(\mu_0+\sigma_0z)|F(x_{n})\right\}\right]\phi(z) dz,
\end{equation}
where $\phi(z)=\exp(-z^2/2)/\sqrt{2\pi}$. The conditional expectation of the equation \eqref{finalpred} is easily evaluated using the Gauss-Hermite quadrature method, see \cite{Ste2002}. Note also that the 
density function \eqref{eq:pred1} can be expressed as:
% \begin{eqnarray}\label{condtionnd}
% g_P(y_n)&=&g(y_n)\frac{c_{}^{(2)}\{F(x_n),G(y_n)\}}{\sigma_0}\times
% \exp\left\{-\frac 1 {2{\sigma}_0^2}\left(\Phi^{-1}\left[C_{2|1}\{G(y_n)|F(x_n)\}\right]-{\mu}_0\right)^2 \right. \nonumber\\
% &&\left. +\frac 1 {2}\Phi^{-1}\left[C_{2|1}\{G(y_n)|F(x_n)\}\right]^2\right\}.
% \end{eqnarray}
\begin{eqnarray}\label{condtionnd}
g_P(y_n)&=&g(y_n)\frac{c_{}^{(2)}\{F(x_n),G(y_n)\}}{\sigma_0}\times
\exp\left\{-\frac 1 {2{\sigma}_0^2}\left(\Phi^{-1}\left[C_{2|1}\{G(y_n)|F(x_n)\}\right]-{\mu}_0\right)^2 +\frac 1 {2}\Phi^{-1}\left[C_{2|1}\{G(y_n)|F(x_n)\}\right]^2\right\}.
\end{eqnarray}
The corresponding quantile function of $g_p$ has a simple form, namely
$$
G_P^{-1}(w)=G^{-1}\left(C_{2|1}^{-1}\left[\Phi\{\mu_0+\sigma_0\Phi^{-1}(w)\}|F(x_n)\cdot
\right]\right),
$$
Thus the median and various quantiles of the conditional distribution of $Y_n$ are easily evaluated.
%\end{proof}

\begin{figure}[H]
\centering
\includegraphics[scale=0.5,width = 16cm, height =10.5cm]{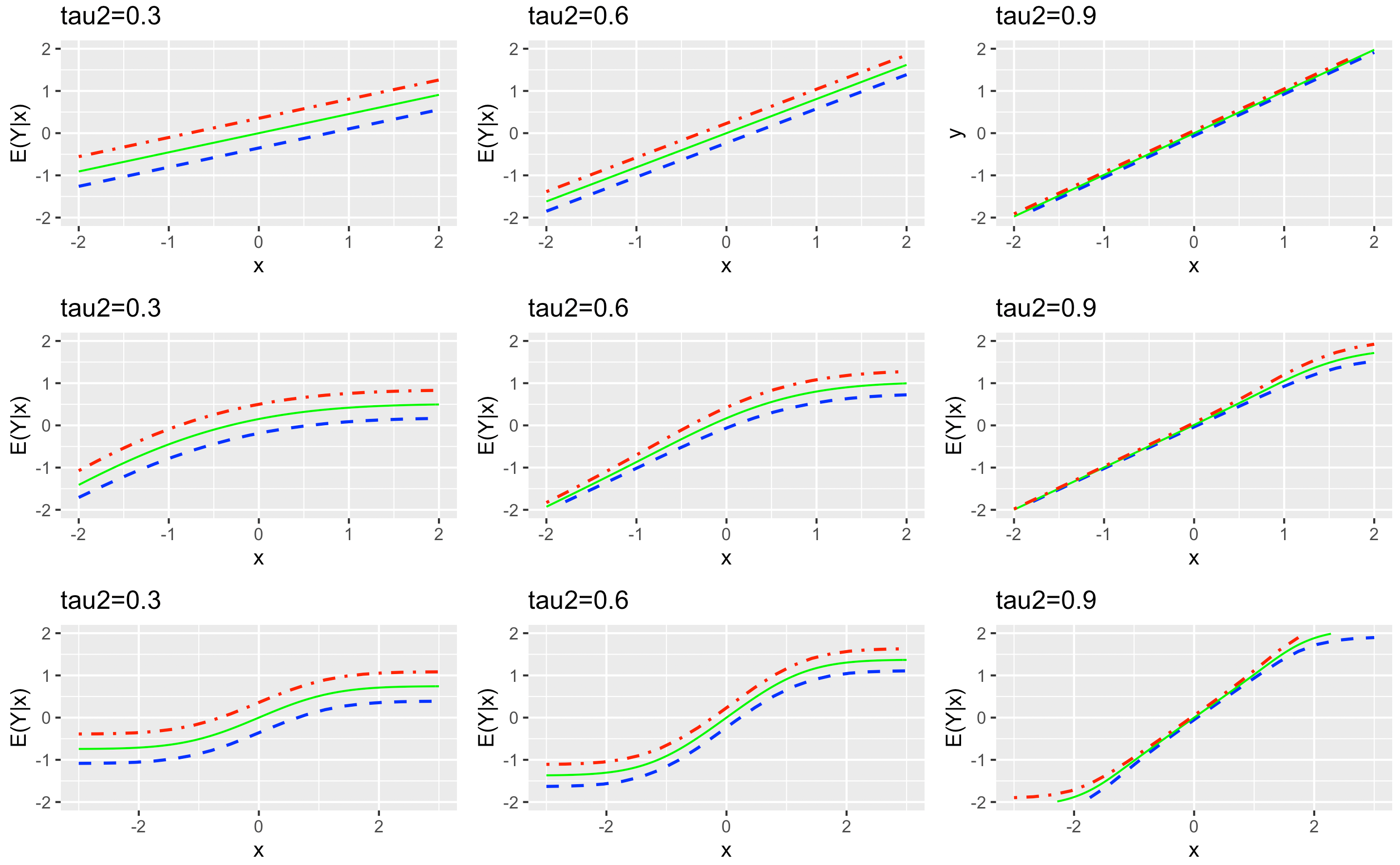}
\caption{Curves of $\mathbb{E}\left(Y|x=x_{n},(x_1,y_1), \ldots, (x_{n-1},y_{n-1})\right)$ with a normal copula (first line), Clayton copula (second line), and Frank copula (third line) for quantiles $q=1/10$ (dotted line), $q=5/10$ (full traint) and $q=9/10$ (dotdash line) of the between cluster effects.}
\label{regcop1}
\end{figure}

When $n$ is large, $\mu_0 \approx \sum_{j=1}^{n-1}\Phi^{-1}(W_j)/(n-1)$ varies between clusters according to a $N(0,\rho_3)$ distribution. This defines the range of possible cluster specific regression curves. 
Using equation \eqref{finalpred}, we construct the prediction curves for three copulas $C_{}^{(2)}$ when the margins $F$ and $G$ are the standard normal distribution. The curves are presented in Figure \ref{regcop1}. They correspond to Kendall correlation coefficient $\tau_2$ equal to $\left(0.3,0.6,0.9\right)$ for each copula $C_{}^{(2)}$. Three copula families $C^{(2)}$ are used: Normal, Frank and Clayton copulas. The dependence parameters are deduced from $\tau_2$ following \citet[pp. 58, 166 and 168]{Joe2014a} and using the function \textit{iTau} of the \textbf{R}-package copula, see \citet{Kojadi10}. We also fix the size $n=21$ and the Kendall tau of $C_{1,1:n}^{(3)}$ is $\tau_3=0.1$ corresponding to $\rho_3=0.16$. Prediction curves are plotted for the quantiles $q\in \{1/10, 5/10, 9/10\}$ of $\sum_{j=1}^{n-1}\Phi^{-1}(W_j)/(n-1)$ whose approximate distribution is a $N(0,\rho_3)$.
The graphs exemplify the impact of the dependency in $C^{(2)}$ on the prediction curves. Indeed, when the copula $C^{(2)}$ is a normal copula, we obtain, in Figure \ref{regcop1}, parallel regression lines corresponding to the mixed linear regression model of \cite{BatHatFul1988}. For Clayton's copula, the regression curves are close for small values of $x$; this might be related to the lower tail dependency of that copula family. When the correlation $\tau_2$ increases, the regression curves tend to a unique straight lines. This agrees with the conditional form of the model given in equations \eqref{decompbat} and \eqref{eq:condmod} where the random part of the model tends to 0 as $\rho_2$ increases to 1. In Figure \ref{regcop1} we used a relatively small value for the error dependence parameter in $C_{1,1:n}^{(3)}$ ; such a small residual dependency is found in many applications such as the analysis presented in Section \ref{section5}.

\section{Copula selection and parameter estimation for a 2-exchangeable model}\label{infstat}

The data set to analyze is $\{(X_{ij},Y_{ij}): i=1,\ldots, m; \ j=1,\ldots,n_i\}$, where index $i$ is for clusters and $j$ is for units within clusters. We assume that the joint distribution of the $2n_i$ variables in cluster $i$ is determined by two marginal distributions $F(x|\alpha)$, $G(y|\beta)$ and by copula density \eqref{eq:ec1} that involves a bivariate copula density, $c^{(2)}(u,v;\delta_2)$ for the marginal relationship between $X$ and $Y$ and two 1-exchangeable copula families with densities, $c_{1,1:n}^{(1)}(u_1,\ldots,u_n;\delta_1)$ and $c_{1,1:n}^{(3)}(w_1,\ldots,w_n;\delta_3)$. The goal of this section is to identify parametric families for the five components of this model and to estimate 
their parameters that are put in vector $\theta=(\alpha, \beta, \delta_1, \delta_2, \delta_3)$.
Following the model building procedure presented in chapter 5 of \citet{Joe2014a} and chapters 7 and 8 of \citet{Caz2019}. The model components are selected sequentially, using techniques that are presented in this section.

\subsection{Determination of the model components}\label{sec:component}
The determination of the marginal distributions is done independently for the two margins. Competing models are compared on the basis of AIC criteria calculated as if the units were independent. Preliminary, or IFM, estimators see \citet[section 5.5]{Joe2014a}, $(\tilde \alpha, \tilde \beta)$ are obtained by maximizing
$$\mathcal{L}_F=\sum_{i=1}^{m} \sum_{j=1}^{n_i} \log \left\{ f(x_{ij}|\alpha) \right\}
\mbox{~~and~~}
\mathcal{L}_G=\sum_{i=1}^{m}\sum_{j=1}^{n_i} \log \left\{ g(y_{ij}|\beta)\right\}\cdot$$
The next step is to calculate the pseudo observations, defined by
\begin{equation} \label{pseudo}
\tilde u_{ij}=F(x_{ij}| \tilde \alpha),\,\,\,\,\,\,\,\,\,\tilde v_{ij}=G(y_{ij}|\tilde \beta)\cdot
\end{equation}

The $N=\sum n_i$ pairs $(\tilde u_{ij},\tilde v_{ij}) $ are then used to select a copula $C^{(2)}$ for the $(X,Y)$ dependency. Note that the bivariate empirical distribution function of the pseudo-observations $(\tilde u_{ij},\tilde v_{ij}) $ is a consistent estimator of $C^{(2)}$ when the number of clusters $m$ goes to infinity provided that the cluster sizes are bounded, $n_i<n_{max}$ for each $i$. The within cluster dependency impacts the variances of the estimators however this is overlooked at this stage and standard methods, proposed in \citet{Joe2014a} and \citet{Caz2019}, are used to select a copula family $C^{(2)}$ for the bivariate sample. Competing models are compared on the basis of their AIC and an IFM estimator $\tilde \delta_2$ of the parameter of the selected copula family is obtained by maximizing
$$\mathcal{L}_2=\sum_{i=1}^{m}\sum_{j=1}^{n_i} \log \left[ c_{ }^{(2)}(\tilde u_{ij},\tilde v_{ij};\delta_2) \right].$$

To assess the within cluster dependency associated with copula families $C_{1,1:n}^{(1)}$ and $C_{1,1:n}^{(3)}$, we use the unit level version of the exchangeable Kendall's tau introduced in \citet{romdhani2014exchangeable}. It is evaluated using the proportion of concordant pairs $\{(x_{ij},x_{i\ell}),(x_{kr},x_{ks})\} $
among the $\sum_{i>k}n_i(n_i-1)n_k(n_k-1)$ possible pairs of ordered observations coming from different clusters. Graphical methods to select a family of exchangeable copulas are proposed in \citet{Riv2016}. Models can also be compared on the basis of their AIC, and an IFM estimator of $\delta_1$ is obtained by maximizing
\begin{equation*} 
\label{eq:loglikc1}
\mathcal{L}_1=\sum_{i=1}^{m} \log \left\{ c_{1,1:n_i}^{(1)}(\tilde u_{i1},\ldots,\tilde u_{in_i};\delta_1) \right\}.
\end{equation*}

The selection of the copula family $C_{1,1:n}^{(3)}$ is based on the pseudo observations $\tilde w_{ij}=C^{(2)}_{2|1}(\tilde v_{ij}|\tilde u_{ij}; \tilde \delta_2)$. The exchangeable Kendall's tau evaluated on
$\{\tilde w_{ij}:\ i=1,\ldots,m; j=1,\ldots,n_i\}$ can be used to assess the within cluster dependency of the residuals. IFM estimator $\tilde \delta_3$ is obtained by maximizing
\begin{equation*} \label{eq:loglikc3}
\mathcal{L}_3=\sum_{i=1}^{m} \log \left\{ c_{1,1:n_i}^{(3)}(\tilde w_{i1}, \dots, \tilde w_{in_i};\delta_3)\right\}.
\end{equation*}

As note in \citet[section 5.5]{Joe2014a} the five estimating equations associated with the IFM estimation of $\theta$ can be combined in a multivariate estimating equation that yield the IFM estimator $\tilde \theta$. The standard asymptotic theory, presented in \citet[pp. 30]{Tsitias2006}, applies. It shows that the joint asymptotic distribution of $\tilde \theta-\theta$, as the number of clusters $m$ goes to infinity, is a centered multivariate normal distribution with a sandwich covariance matrix. As expected, this sandwich variance estimator accounts for the within cluster dependence.

\subsection{Maximum likelihood estimation of the parameters}

Once parametric families for the five model components have been identified, optimal estimators of the parameters are obtained by maximum likekihood. This section discusses the properties of the maximum likelihood estimator for the parameter vector $\theta$. 

The log-likelihood for $\theta$ is equal to
\begin{eqnarray*}
\mathcal{L}(\theta)&=&\sum_{i=1}^{m} \sum_{j=1}^{n_i} \log \left\{ f(x_{ij}|\alpha) \right\}+ \sum_{i=1}^{m}\sum_{j=1}^{n_i} \log \left[ c_{ }^{(2)}\left\{F(x_{ij} |\alpha),G(y_{ij}|\beta);\delta_2\right\} \right] \\
&+& \sum_{i=1}^{m}\sum_{j=1}^{n_i} \log \left\{ g(y_{ij}|\beta)\right\} +
\sum_{i=1}^{m} \log \left[ c_{1,1:n_i}^{(1)}\left\{F(x_{i1}|\alpha),\ldots,F(x_{in_i}|\alpha);\delta_1\right\} \right] \nonumber\\
&+& 
\sum_{i=1}^{m} \log \left[ c_{1,1:n_i}^{(3)}\left[C_{2|1}\left\{G(y_{i1}|\beta)|F(x_{i1}|\alpha)\right\},\ldots,C_{2|1}\left\{G(y_{in_i}|\beta)|F(x_{in_i}|\alpha)\right\};\delta_3\right]\right]\nonumber\cdot
\end{eqnarray*}
This function is easily maximized once parametric families for the two margins in the model and the three copulas are selected. This yields $\hat \theta$ the maximum likelihood estimator of the parameter vector. This maximization is carried out using an optimiser such the R-function \textit{optim} or \textit{nlminb}. Minus the hessian of $\mathcal{L}(\theta)$, evaluated at $\hat \theta$, is the observed Fisher information for the model. Its inverse is the asymptotic covariance matrix of $\hat \theta -\theta$. It can be used to calculate standard error estimates for all the parameters that have been estimated. At this stage, likelihood ratio tests comparing nested candidate parametric families for $C^{(2)}$ can also be carried out to validate the IFM model selection step that ignored the within cluster dependency.

When the margins $F(x)$, $G(y)$ and the copula $C^{(2)}$ are normal, the likelihood can be split into a marginal likelihood for the parameters of $F(x)$ and $C_{1,1:n}^{(1)}$ times a conditional likelihood for the regression parameters $(\beta_0,\beta_1,\sigma_e^2)$ and the parameter for $C_{1,1:n}^{(3)}$, see equation \eqref{eq:condmod}. The standard normal mixed linear model of \cite{BatHatFul1988} falls into that category as its parameters are estimated using a conditional likelihood. In general the parameters are intertwined in a complicated way and their estimation relies the log-likelihood $\mathcal{L}(\theta)$.

% \begin{proposition}\textbf{Generalization of the IFM method}
% To fit a 2-changeable copula model, we fit the marginals of the laws, and then successfully fit the copulas $C_{1,1:n}^{(1)}$, $C_{}^{(2)}$ and $C_{1,1:n}^{(3)}$. In particular, if the copulas $C_{1,1:n}^{(1)}$ and $C_{1,1:n}^{(3)}$ are independence copulas, we end up with the IFM method, see \cite{joe1996}.
% \end{proposition}

% \begin{proof}
% Under the assumptions defined by the \ref{propasymptotic} proposition, the method leads to the same results as the maximum likelihood method, but with large variance for the estimators. If the copulas $C_{1,1:n}^{(3)}$ and $C_{1,1:n}^{(3)}$ are independence copulas then their density is 1 and we just adjust the marginals and the copula $c_{}^{(2)}$.
% \end{proof}

% The calculations for obtaining these results can be found in the supplementary material. This result is a special case of the comparison of the variance-covariance matrix of the estimators of the 2-exchangeable copula model when all copulas are normal. We study in the following the cases where the exchangeable copulas are normal and the copula $C^{(2)}$ varies.

\subsection{A Monte Carlo investigation of the sampling properties of  IFM and  ML estimators} \label{simulsec}

For a 2-exchangeable copula model with parameters $\theta$, $\hat{\theta}$ and $\tilde{\theta}$ are the maximum likelihood and the IFM estimators. This section investigates their sampling properties when $m$ is finite. This is done using Monte-Carlo simulations where the expectation and the variance of an estimator $\hat{\psi}$ are approximated by
\begin{equation*}\label{estimation_montecarlo}
\mathbb{E}_B(\hat{\psi}) = \frac{1}{B}\sum_{b=1}^B \hat \psi_b,\,\,\,\,\,\,
\mathbb{V}_B(\hat{\psi}) = \frac{1}{B-1}\sum_{i=1}^B\left\{\hat \psi_b-\mathbb{E}_B(\hat{\psi}) \right\}^2,
\end{equation*}
where $b$ indexes the estimates obtained in the $B=1000$ Monte-Carlo simulations. The expectations and the variances of IFM estimators $\tilde \psi$ are evaluated in a similar way.

Throughout this Monte Carlo study of the copula 2-exchangeable model, the margins $F$ and $G$ are normal distributions with mean $\mu_1=\mu_2=0$ and variance $\sigma_1^2=\sigma_2^2=1$. The exchangeable copulas $C_{1,1:n}^{(1)}$ and $C_{1,1:n}^{(3)}$ belong to the normal family with correlation $\rho_1=0.31$ and $\rho_3=0.16$ respectively, corresponding to Kendall's tau of 0.2 and 0.1. Such small levels of within cluster association are often found in applications. In the study correlations are parameterized in terms of their logit,
\begin{equation} \label{eq:logit}
  \eta= \log \left( \frac{\rho}{1-\rho}\right).  
\end{equation} 
Three copulas $C^{(2)}$ are investigated: the normal copula, the Clayton copula and a two-parameter  Khoudraji copula \citep{genest1998discussion} that features an asymmetric relationship between $U$ and $V$.  It is defined by
\begin{equation}\label{eq:khou}
C_{}^{(2)}(u,v;\rho,\kappa)=u^{1-\kappa}C_{\rho}(u^{\kappa},v),\,\,\,\,u,v\in [0,1],
\end{equation}
where $C_\rho$ is a normal copula with correlation $\rho \in (0,1)$ and $\kappa \in (0,1)$ is the asymmetry parameter. In the simulations $\kappa$ is parameterized in terms of its logit, see \eqref{eq:logit}. Two values for the  Kendall's tau $\tau_2$ of copula $C_{}^{(2)}$, 0.4 and 0.6, are considered.  Two sample sizes, $m=10,50$, are investigated; their corresponding cluster sizes are $n=30$ and $n=18$ respectively. Tables \ref{table:Resultat de la simulation obtenuscenario1} to \ref{simul_clay_glob} present the expectations and the variances of the estimators of the copula parameters.

%\spacingset{1.2}
\begin{table}[H]
\caption{Expectations of the estimators and their with variances multiplied by 10, in parenthesis, when $C_{}^{(2)}$ is a normal copula.}
\centering
%\tiny 
\begin{tabular}{llcccc}
\toprule
$\tau_2(\eta_2)\,\,\,\,$ & $m$ & Method & $\eta_1=-0.80$ & $\eta_2$ & $\eta_3=-1.69$\\ 
\toprule
%%1%%%%%%%%%%%%%%%% 
 \multirow{1}{2em}{0.4(0.35)} & \multirow{2}{2em}{10} & MV &  -0.96(1.90) & 0.35(0.37) & -1.81(2.29) \\
%& 
& & IFM & -1.00(2.20) & 0.34(0.62) & -1.89(2.41)\\ %%%FAIT
 \cline{2-6}
 & \multirow{2}{2em}{50} & MV & -0.85(0.58) & 0.35(0.12) & -1.74(0.75) \\
 %& 
 & & IFM & -0.86(0.63) & 0.35(0.16) & -1.76(0.80)\\%%%%%%FAIT%%%%%%%%%%%%%
\toprule
%%1%%%%%%%%%%%%%%%% 
 \multirow{1}{2em}{0.6(1.44)} & \multirow{2}{2em}{10} & MV & -0.90(1.97) & 1.45(0.31) & -1.83(2.24) \\
%& 
& & IFM & -0.95(2.30) & 1.44(0.41) & -1.93(2.50)\\ %%%FAIT
 \cline{2-6}
 & \multirow{2}{2em}{50} & MV & -0.84(0.53) & 1.44(0.09) & -1.73(0.70) \\
 %& 
 & & IFM & -0.85(0.60) & 1.44(0.11) & -1.75(0.76)\\%%%%%%FAIT%%%%%%%%%%%%%
\toprule
\end{tabular}
\label{table:Resultat de la simulation obtenuscenario1}
\end{table}

\spacingset{1.9}
\begin{table}[H]
\caption{Expectations of the estimators and their with variances multiplied by 10, in parenthesis,  when $C_{}^{(2)}$ is Clayton copula.}
\centering 
%\tiny 
\begin{tabular}{llcccc} %\toprule
\toprule
$\tau_2 (\delta_2)\,\,\,\,$ & $m$ &  Method & $\eta_1=-0.80$ & $\delta_2$ & $\eta_3=-1.69$\\ 
\toprule
 \multirow{1}{2em}{0.4(1.33)} & \multirow{2}{2em}{10} & MV &  -0.88(1.63) & 1.35(0.66) & -1.82(2.36) \\
& &  IFM &  -0.98(2.18) & 1.33(0.84) & -1.91(2.68)\\ %%%FAIT 
 \cline{2-6}
 & \multirow{2}{2em}{50} &  MV & -0.83(0.46) & 1.33(0.15) & -1.73(0.67) \\
 %&  
 & & IFM & -0.85(0.60) & 1.32(0.19) & -1.75(0.73)\\%%%%%%FAIT%%%%%%%%%%%%%
\toprule
\multirow{1}{2em}{0.6(3)} & \multirow{2}{2em}{10} & MV &  -0.88(1.48) & 3.02(2.39) & -1.80(2.48)\\
& &  IFM &  -0.98(2.11) & 2.96(3.13) & -1.91(2.56)\\ %%%FAIT 
 \cline{2-6}
 & \multirow{2}{2em}{50} &  MV & -0.82(0.45) & 3.01(0.61) & -1.73(0.71) \\
 %& 
 & & IFM & -0.86(0.59) & 2.99(0.77) & -1.75(0.77)\\%%%%%%FAIT%%%%%%%%%%%%%
\toprule
\end{tabular}
\label{simul_clay_glob}
\end{table}

 %\spacingset{1.2}

%LEGENDE : https://www.tandfonline.com/action/authorSubmission?show=instructions&journalCode=uasa20
\begin{table}[H]
\caption{Expectations of the estimators and their with variances multiplied by 10, in parenthesis,  when $C_{}^{(2)}$ is Khoudraji copula.}
\centering 
\begin{tabular}{llccccc} 
\toprule
$\tau_2\,\,\,\,$ & $m$ & Method  &$\eta_1=-0.80$ & $\eta_{\rho}=0.75$ & $\eta_{\kappa}=1.52$ & $\eta_3=-1.69$\\ 
\toprule  
\multirow{1}{2em}{0.4} & \multirow{2}{2em}{10}  & 
        MV & -0.90(1.83) & 0.80(0.55) &  1.51(2.42) & -1.85(2.13) \\
& &  IFM & -0.95(2.08)& 0.75(0.65) & 1.49(3.25) &  -1.89(2.49)\\ 
 \cline{2-7}
%%1%%%%%%%%%%%%%%%% 
  & \multirow{2}{2em}{50} & MV  & -0.82(0.51) & 0.77(0.12) & 1.51(0.98) & -1.73(0.63) \\
  %&
  & &  IFM  & -0.83(0.58) & 0.76(0.18) & 1.51(1.44) & -1.74(0.79)\\%%%FAIT%
  \toprule %%%%FAIT 
$\tau_2\,\,\,\,$ & $m$ & Method  &$\eta_1=-0.80$ & $\eta_{\rho}=1.45$ & $\eta_{\kappa}=3.48$ & $\eta_3=-1.69$\\
\toprule
\multirow{1}{2em}{0.6} & \multirow{2}{2em}{10}  & 
        MV & -0.86(1.77) & 1.49(0.36) & 3.39(5.30)  & -1.81(2.20) \\
& &  IFM &  -0.94(2.09) & 1.45(0.38) & 3.62(6.34) & -1.88(2.74)\\ 
 \cline{2-7}
%%1%%%%%%%%%%%%%%%%
  & \multirow{2}{2em}{50} & MV  & -0.81(0.49) & 1.47(0.11) & 3.48(2.14) & -1.77(0.73)\\
  %&
  & &  IFM  & -0.83(0.63) & 1.47(0.14) & 3.50(2.79) & -1.79(0.86)\\%%%FAIT%
  \toprule %%%%FAIT
\end{tabular}
\label{table:essai}
\end{table}
 \spacingset{1.9}
In Tables \ref{table:Resultat de la simulation obtenuscenario1} %, \ref{simul_clay_glob}  
to \ref{table:essai}, all estimators have negligible biases. The discussion focuses on variances. As expected the strength of the $U$-$V$ association in $C^{(2)}$ does not impact the precision of the two estimators for $\eta_1$. The loss of precision for the IFM estimators is larger for the parameters of copula $C^{(2)}$ than for the parameters of the other 2 copula families. The efficiency of the maximum likelihood estimator is larger at $m=10$ than at $m=50$. The Supplementary Material provides additional simulations for unequal sample sizes $\{n_i\}$ within clusters.  Unequal sample sizes are associated to a small loss of precision for all estimators, especially at $m=10$. This loss is, in general, more important for IFM estimators than for maximum likelihood estimators. Overall the two estimation methods give similar results;  this supports the proposal of Section \ref{infstat} to use IFM estimators to select the components of the proposed copula model. The detailed simulation results, including a presentation of the sampling properties of the estimators for $(\mu_k,\sigma_k)$ for 
$k=1,2$, are presented in the Supplementary Material.

 % \spacingset{1.5}

\section{Modeling math grades with a 2-exchangeable copula model}\label{section5}
This section revisits a data set discussed in \cite{Gold2011}. It concerns math grades in fourth and seventh year measured on $N=\sum n_i=728$ students in $m=48$ primary schools. The cluster sample sizes $n_i$ vary between 4 and 40. The fourth year mark ($X$) and the seventh year mark ($Y$) vary between 0 and 40. We map them to the (0,1) interval using the transform $M \rightarrow (M+1/2)/41$. To break ties in the grades a small random perturbation was added to each one.
\begin{figure}[H]
\centering
\includegraphics[height=7.0cm,width=11cm,scale=0.7]{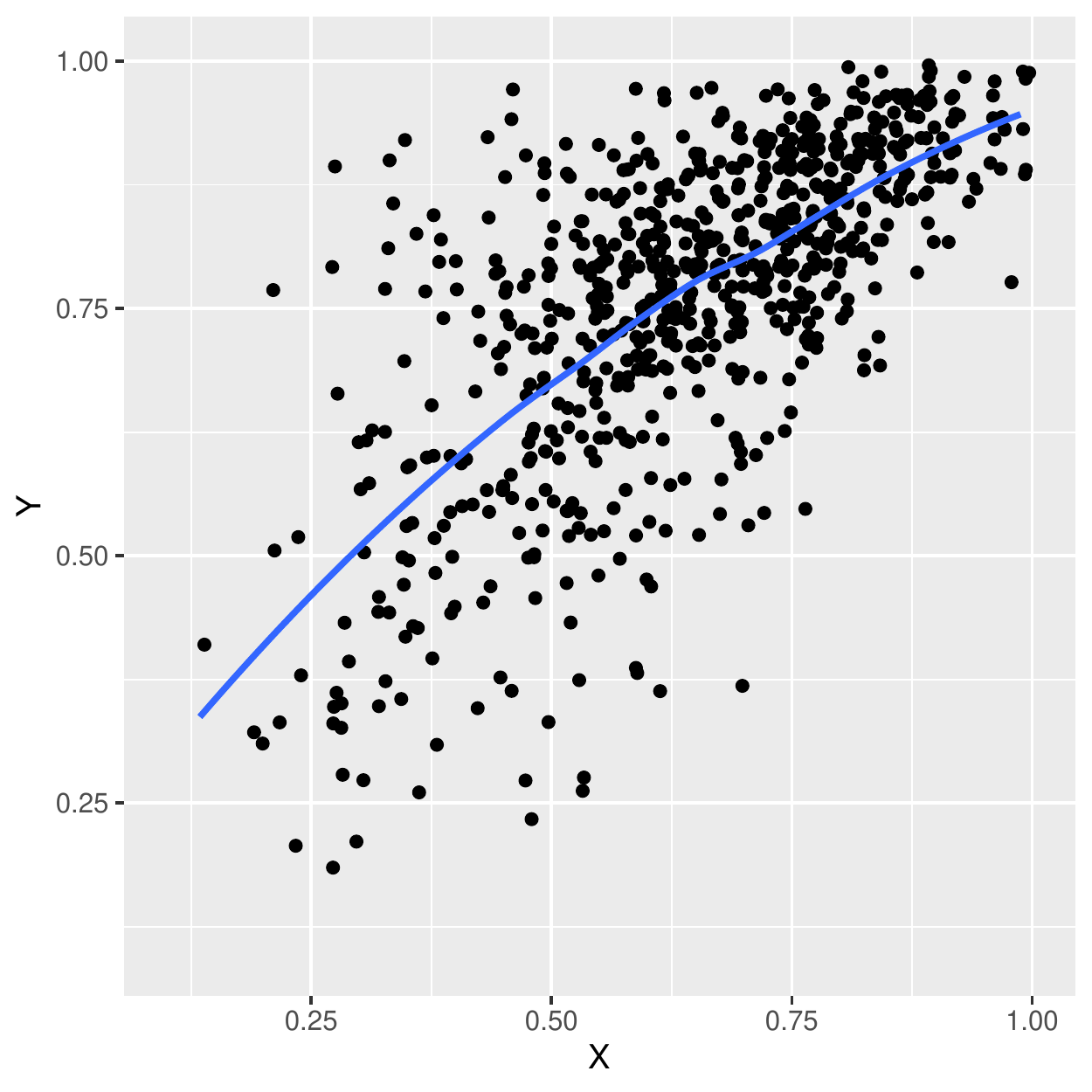}
\caption{Scatter plot of $Y$ versus $X$ with a smooth.}
\label{fig:smooth}
\end{figure}
One objective of the analysis presented here is to construct predictive models for $Y$ given $X$ whose support is $(0,1)$. Another objective is to investigate whether the 2-exchangeable copula model can capture the non-linearity seen in Figure \ref{fig:smooth} that gives a scatter plot of the $N=728$ data points and a smooth. The goal is to contrast an analysis carried out with copulas to the one reported in \cite{Gold2011} that is based on standard normal linear mixed models.

\subsection{Selection of the marginal distributions for $X$ and $Y$}
The candidates families for $F$ and $G$ are the beta (denoted $\mathcal{B}$) and the generalized beta (denoted $\mathcal{G}\mathcal{B}3$) distributions, see \cite{Coc2018}. Indeed, by the construction of histogram of the distribution, we arrive at an asymmetric law. If $X$ has a $\mathcal{B}(\alpha, \beta)$ distribution then $Y=X/\{\lambda+(1-\lambda)X\}$ has for $\lambda \in (0,1)$ a $\mathcal{G}\mathcal{B}3(\alpha, \beta, \lambda)$ distribution whose density is given by
\begin{equation*}
\frac{\lambda^\alpha\Gamma(\alpha+\beta)}{\Gamma(\alpha)\Gamma(\beta)}\frac{y^{\alpha-1}(1-y)^{\beta-1}}{\left\{1-(1-\lambda)y \right\}^{\alpha+\beta}},\,\,\,\,\,\, 0\leq y \leq 1.
\end{equation*}
Table \ref{indemargi} compares the fit of these two distributions to the two margins. As stated in Section \ref{sec:component}, this preliminary analysis does not account for the within cluster dependency. Thus the $pse$, for pseudo standard error, ignores the within classroom dependency.
\begin{table}[H] 
\begin{center}
\caption{Fit of the $\mathcal{B}$ and of the $\mathcal{G}\mathcal{B}3$ distributions to the two margins.} 
\label{indemargi}
\begin{tabular}{ccccc}
\toprule
Variable & Model & $\tilde \theta$ & $pse$& AIC \\
\hline 
$X$ & $\mathcal{B}$ & (4.280,2.332) & (0.222,0.115) & -542.88 \\
 & $\mathcal{G}\mathcal{B}3$ & (4.280,2.330,1) & (0.222,0.115,NA) & -540.80\\
\hline 
$Y$ & $\mathcal{B}$ & 5.24,1.79 & (0.28,0.08) & -789.57\\
 & $\mathcal{G}\mathcal{B}3$ & (2.616,2.319,0.29) & (0.303,0.241,0.069) & -826.96\\
\toprule
\end{tabular}
\end{center} 
\end{table}
The best fitting models are respectively the beta end the generalized beta for $X$ and $Y$. 
%The histograms of the two variables and the fitted densities are given in Figure \ref{fig:my_label2}. This illustrates the increase in both, skewness and kurtosis, of the marks in the seventh grade. A three parameter beta distribution is needed to capture this feature.
%\begin{figure}[H]
% \centering
 %\includegraphics[scale=0.7,height=8cm,width=13cm]{Ajustx-y.pdf}
%\caption{Histograms for $X$ and $Y$ with the density of the best \texttt{\textit{ AIC}} models superimposed}
 %\label{fig:my_label2}
%\end{figure}

\subsection{Selection of the bivariate copula $C^{(2)}$}
The fist step is to calculate the pseudo observations $\tilde{u}_{ij}$ et $\tilde{v}_{ij}$ for $j=1,\ldots,n_i$ and $i=1,\ldots,m$ defined in \eqref{pseudo}. The Kendall's tau is 0.49 ($pse=0.03$) so there is a relatively strong association between the two variables. Following \cite[chap. 1]{Joe2014a}, Kendall's tau is calculated for the sub-samples in the 4 quadrants of the unit square. This reveals a stronger association for large grades than for smaller ones. Also a 0.1 difference between the Kendall's tau for the upper left and lower quadrant suggests that some of the asymmetry seen in the Figure \ref{fig:smooth} is left once the margins' effect has been factored out. 

Several copulas were fitted to this bivariate sample using the functions \texttt{BiCopEst} in the R package \texttt{VineCopula} and \texttt{fitCopula} in \texttt{copula}. To capture the asymmetry in the data we used to Khoudraji device, see \eqref{eq:khou} to create asymmetric alternatives. The best fitting copula in Table \ref{tabsymetri} is the survival Khoudraji normal copula given by
$$
C^{(2)}(u,v|\rho_2,\kappa_1,\kappa_2)=u+v-1+(1-u)^{1-\kappa_1}(1-v)^{1-\kappa_2}C_{\rho_2}\{
(1-u)^{\kappa_1},(1-v)^{\kappa_2}\},
$$
where $C_{\rho_2}$ is the bivariate normal copula with correlation $\rho_2$. The density of a copula in this three parameter family can be evaluated using functions of \texttt{copula}. The conditional distribution, $ w=\partial C^{(2)}(u,v| \rho_2,\kappa_1,\kappa_2)/\partial u$ has the following explicit form.
\begin{eqnarray}\label{pseu1}
w&=&1-(1-{\kappa}_1)\left(1-u\right)^{-{\kappa}_1}\left(1-v\right)^{1-{\kappa}_2}C_{\rho_2}\left\{ \left(1-u\right)^{{\kappa}_1},\left(1-v\right)^{{\kappa}_2} \right\}-{\kappa}_1(1-v)^{1-{\kappa}_2}\Phi\left[\frac{\Phi^{-1}\left\{(1-v)^{{\kappa}_2}\right\}-\rho_2\Phi^{-1}\left\{(1- u)^{{\kappa}_1}\right\}}{\sqrt{1-{\rho_2}^2}}\right]\cdot
\end{eqnarray}
In Table \ref{tabsymetri}, Survival Khoudraji-Normal2 refers to a two parameter version of this copula obtained by setting $\kappa_2=1$.

\begin{table}[H]
\centering
\caption{Paramater estimates, their pseudo standard errors $pse$ and the AIC for several copulas for the $(u,v)$ relationship.}
\begin{tabular}{lccc}
\toprule
 Copula $C^{(2)}$ & $\tilde \theta$ & $pse$ & AIC \\
\toprule 
Normal & $0.682$ & $0.016$ & -454.32 \\ 
Survival-Gumbel & $1.892$ & $0.057$ & -459.5 \\ 
Survival Khoudraji-Normal & $(0.791,0.822,0.960)$ & $(0.023,0.043,0.029)$ & -474.10 \\
Survival Khoudraji-Normal2 & $(0.762,0.837)$ & $(0.021,0.045)$ & -468.64 \\
\toprule 
\end{tabular}
\label{tabsymetri}
\end{table}

\subsection{Selection of the exchangeable copula families $C_{1,1:n}^{(1)}$ and $C_{1,1:n}^{(3)}$ }

The exchangeable Kendall's tau for $X$ and $Y$ are respectively 0.046 ($se=.017$) and 0.095 ($se=0.035$) showing a stronger school effect in the seventh year. The fit of several families of copulas for the joint distribution of $\{\tilde u_{ij}\} $ and $\{\tilde w_{ij}\}$, evaluated using \eqref{pseu1}, are compared by maximizing the pseudo log-likelihoods $\mathcal{L}_1$ and $\mathcal{L}_3$. 
The results are reported in Table \ref{exchc2}. The normal family is the best choice for both $C_{1,1:n}^{(1)}$ and $C_{1,1:n}^{(3)}$.
\begin{table}[H] 
\centering
\caption{Parameter estimates, their pseudo standard errors $pse$, and the AIC for three families for $C_{1,1:n}^{(1)}$ and $C_{1,1:n}^{(3)}$.}
\begin{tabular}{lcccccc}
& \multicolumn{3}{c}{$C_{1,1:n}^{(1)}$} & \multicolumn{3}{c}{$C_{1,1:n}^{(3)}$} \\
\toprule 
Family & $\tilde \theta$ & $pse$ & AIC & $\tilde \theta$ & $pse$ & AIC\\
\toprule 
Frank & 0.251 & 0.138 & -5.69& 0.935 & 0.179 & -69.76\\
Gumbel & 1.040 & 0.022 & -4.64 & 1.109 & 0.024 & -66017\\
Normal & 0.064 & 0.025 & -12.76 & 0.167 & 0.035 & -80.03\\
\toprule
\end{tabular}
\label{exchc2}
\end{table}

To complete the analysis we carry out a full maximum likelihood estimation of the 10 parameters for the components of the proposed model. The parameter estimate for $\kappa_2$ is very close to 1. We first carry out a likelihood ratio test for $H_0:\ \kappa_2=1$ using the full likelihood. This gives $\chi^2_{1,obs}=8.2$ for a p-value of 0.4\%. Thus the full model, with 10 parameters, is definitive ; the estimates and standard errors ($se$) for the parameters are reported in Table \ref{estim_gobale}.

\begin{figure}[H]
 \centering
 \includegraphics[scale=0.7]{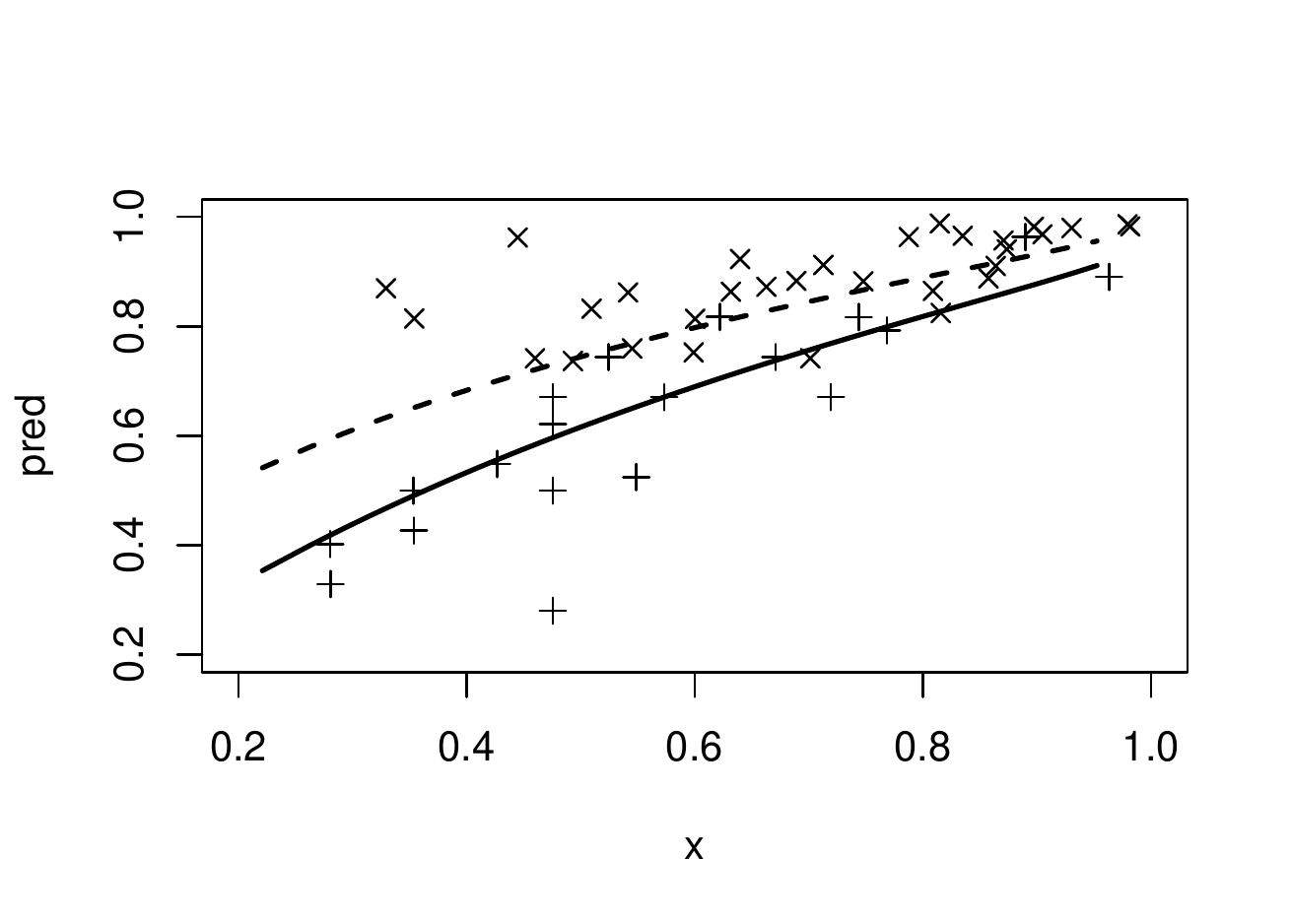}
 \caption{Scatter plots and regression curves for Schools 1 ($+$ character and full line) and 30 ($\times$ character and dashed line).}
 \label{fig:ecol1_30}
\end{figure}

\begin{table}[H]
\centering
\caption{Maximum likelihood estimators of the parameters for the full model.}
\begin{tabular}{lcc}
 \toprule 
Component & $\hat \theta$ & $se$ \\
\toprule
\ $F$ ($\mathcal{B}$) & $(\hat{\alpha}_1,\hat{\beta}_1)=(4.271,2.359)$ & $(0.235,0.124)$\\
  $G$ ($\mathcal{GB}3$) & $(\hat{\alpha}_2,\hat{\beta}_2,\hat{\lambda})=(2.457,2.470,0.248)$ & $(0.245,0.254,0.052)$\\
$C_{1,1:n}^{(1)}$ (Normal) & $\hat{\rho}_1=0.063$ & $0.026$ \\
$C^{(2)}$ (Survival Khoudraji-Normal ) & $(\hat{\rho},\hat{\kappa}_1,\hat{\kappa}_2)=(0.795,0.822,0.959)$ & $(0.024,0.046,0.029)$ \\
$C_{1,1:n}^{(3)}$ (Normal) & $\hat{\rho}_3=0.161$ & $ 0.040$ \\
\toprule
\end{tabular}
\label{estim_gobale}
\end{table}
The fit of the final model summarized in Table \ref{estim_gobale} is illustrated using two schools, numbered 1, with $n_1=19$, and 30, with $n_{30}=31$. Their respective values of $\mu_0$, see \eqref{condimom}, are  $-0.464$ and $0.810$; this means that, for the same math4 mark, the math7 grade in School 30 are higher than in School 1. This can be seen in Figure \ref{fig:ecol1_30} that gives the two regression curves, constructed using \eqref{finalpred}, that use the Gauss-Hermite quadrature method to approximate the normal integrals for each $x$-value. 

Conditional residuals for the fitted copula models can be defined as $y_{ij}-\hat y_{ij} $, where $\hat y_{ij}$ is evaluated as the predicted value at $x_{ij}$, using  \eqref{finalpred}, where the values of $\mu_0$ and $\sigma_0$, see \eqref{condimom}, are those for school $i$. The conditional residuals of the copula model ($CM$) can be compared to those of mixed linear models. The first one, $ML1$, has a random school intercept while the second one, $ML2$, has possibly dependent random slope and intercept. The mean squared errors and inter quartile ranges (IQR) of the residuals for $CM$, $ML1$, and $ ML2$,  are (0.0107, 0.0112, 0.0101) and (0.1082, 0.1186, 0.1098) respectively. Thus, in agreement with the analysis reported in Table \ref{tabsymetri}, the fit of $ML1$ is poor. The fits of $CM$ and $ML2$  are very similar as the latter captures the between school change in slope than can be seen in Figure \ref{fig:ecol1_30}. $CM$ has a larger residual MSE; however it has a smaller residual IQR and it gives a smaller absolute residual than  $ML2$  for $53\%$ of the data points.

 \begin{figure}[H]
 \centering
 \includegraphics[scale=0.7]{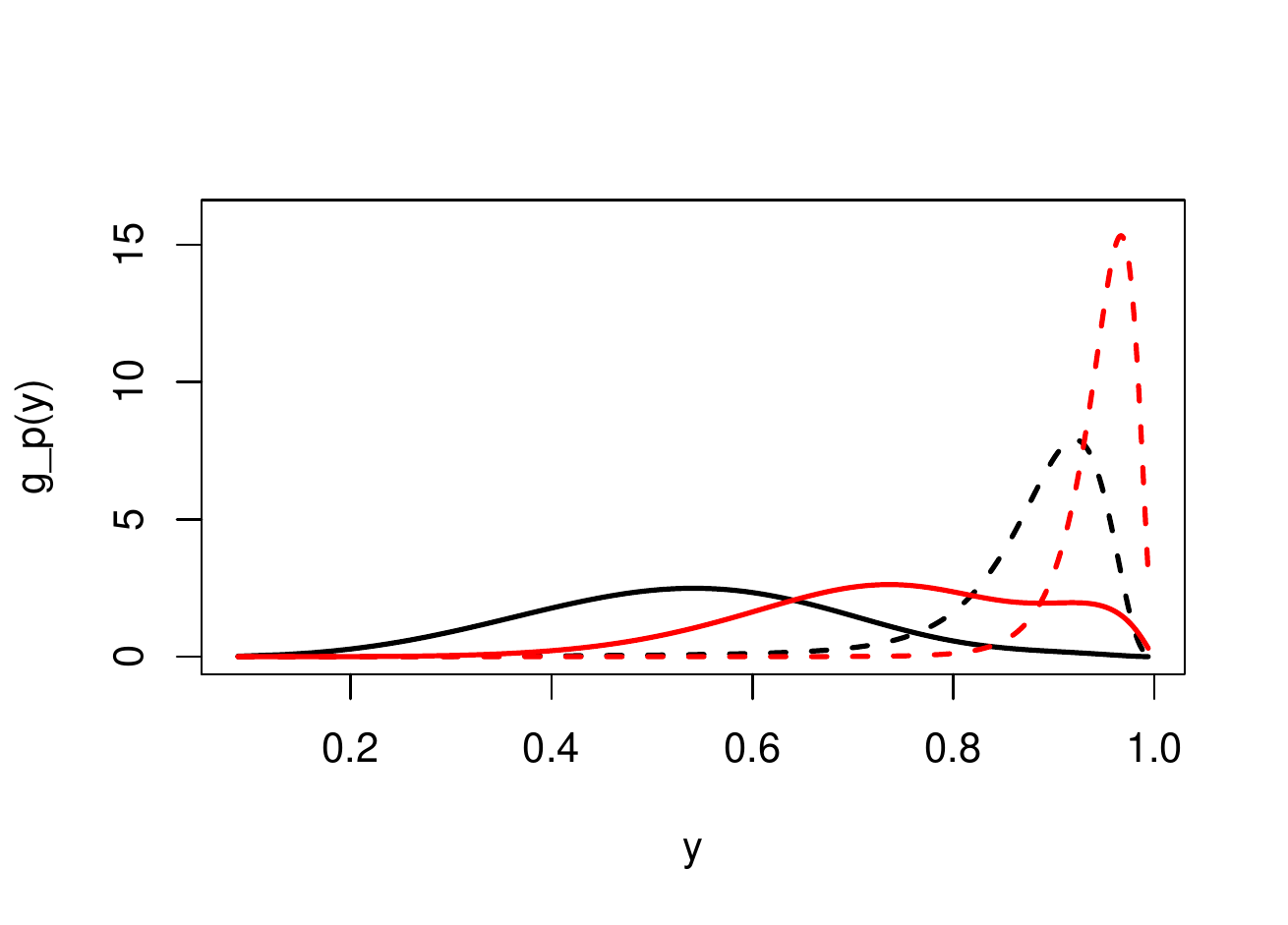}
 \caption{Densities for the predicted $Y$ values for $X=0.4$ (continuous lines) and $X=0.9$ (dashed lines) for Schools 1, in black, and 30, in red.}
 \label{fig:ecol21_30}
\end{figure}

The key advantage of the 2-exchangeable copula model is that it allows prediction intervals for $Y$ to depend on both, the School and the $X$-value. This is illustrated in Figure \ref{fig:ecol21_30} that gives the predictive densities for mark $Y$, given by formula\eqref{condtionnd}, for $X=0.4$ and $X=0.9$ in Schools 1 and 30.  The variability of $Y$ is  larger at $X=0.4$ and in School 1. This can be seen by looking at the corresponding 95\% prediction intervals for $Y$ that are given in Table \ref{fig:pred_int}. 

\begin{table}[H] 
\begin{center}
\caption{95\% prediction intervals for $X=0.4$ and $X=0.9$ in Schools 1 and 30. } 
\label{fig:pred_int}
\begin{tabular}{ccc}

\hline
School &$X=0.4$ & $X=0.9$ \\
\hline 
1 & (0.23, 0.89) & (0.66,0.97) \\
30 & (0.44,0.97)  & (0.86,0.99) \\
\hline
\end{tabular}
\end{center} 
\end{table}

\section{Conclusion }\textbf{}\
The 2-exchangeable copula model proposed in this work provides flexible methods to predict variable $Y$  knowning $X$ in a hierarchical data set. A key feature of the proposed methodology highlighted in Section \ref{infstat} is the flexibility of the predictive densities for $Y$ given $X$. Its shape and its support can depend on the known $X$ value for $Y$ and on the cluster. An outstanding problem is whether the proposed model can be generalized to two or more continuous explanatory variables. The key to such a generalization is the availability of flexible $(d-1)$-exchangeable families $C_{d-1,1:n}^{(1)}$ for the joint distribution of the explanatory variables for all the unit in a cluster. An elliptical copula, with a correlation matrix given by \eqref{eq1} could be used. This specifies partially the copula $C^{(2)}$ for the joint distribution of the $X$ and the $Y$ variables on a unit. A vine decomposition could possibly be used to complete the specification of $C^{(2)}$. These problems will be the object of future investigations.

\section*{Supplementary Material}
The supplementary material contains the proof of Proposition and additional results of the of the Monte Carlo simulation study.

\section*{Acknowledgments and Miscellaneous}

We would like to thank Étienne Marceau for his critical reading of a previous version of this work.

\section*{Funding}
The support of the Natural Sciences and Engineering Research Council of Canada is gratefully acknowledged.

\section*{Disclosure Statement}
The author(s) declared no potential conflicts of interest with respect to the research, authorship, and/or publication of this article.

 \section*{Data availability}
The data is available upon request.

\bibliography{Bibliography_TAG.bbl}

% \newpage
% \appendix
% \LARGE{\textbf{Appendix}}

%%%5MATERIEL SUPPLEMENTAIùrE

\end{document}

\frontmatter

\def\spacingset#1{\renewcommand{\baselinestretch}%
{#1}\small} \spacingset{1.9}

%%%%%%%%%%%%%%%%%%%%%%%%%%%%%%%%%%%%%%%%%%%%%%%%%%%%%%%%%%%%%%%%%%%%%%%%%%%%%%

%\newpage
\if0\blind
{
 \bigskip
 \bigskip
 \bigskip
 \begin{center}
 {\LARGE\bf Supplementary material for ‘‘A new copula regression model for hierarchical data."}
\end{center}
 \medskip
} \fi

\spacingset{1.9} % DON'T change the spacing!
\vspace{0.2cm}

This document contains the proof of corollary in the paper and the detailed results for the simulation study presented in Section 4.3. In addition, we also present the R programs associated with the case study of the 2-exchangeable copula model in section 5.

\section{Proof of Proposition 1} \label{propdemo}
Let $\left\{\bm{Z}_{1}, \ldots, \bm{Z}_{n}\right\}$, a set of $n$ random vectors each of which is of dimension $d$ and verifying the $d$-exchangeability condition. The covariance between two sub-vectors $\bm{Z}_j$ and $\bm{Z}_k$, $j, k=1, \ldots,n$ does not depend on $j$ and $k$; it is denoted $\Sigma_b$. In addition the variances of the sub-vectors $\bm{Z}_{j}, j=1, \ldots,n$ are all equal to a covariance matrix denoted $A=\Sigma_w + \Sigma_b$. Thus, the covariance matrix of $\left\{\bm{Z}_{1}, \bm{Z}_{2},\ldots, \bm{Z}_{n}\right\}$ has the form given in proposition 1. To prove that both $\Sigma_w$ and $\Sigma_b$ are semi positive definite for Spearman and Pearson correlation matrix, observe that
for any vector $ u \in \mathbb{R}^{d}$, $\{u^\top\bm{Z}_{j}\}$, or a linear combination of the marginal ranks, is a 1-exchangeable vector and the correlation matrix of $(u^\top\bm{Z}_{1}, u^\top\bm{Z}_{2})$ is given by:
\begin{equation*} \label{p1}
\left(\begin{array}{cc}
	u^\top\left(\Sigma_w + \Sigma_b\right)u & u^\top\Sigma_b u \\ 
	u^\top\Sigma_b u & u^\top\left(\Sigma_w + \Sigma_b\right)u
	\end{array}\right) 
\end{equation*}
According to \citet[pp. 41]{MaiSher2012} 1-exchangeability implies that the covariance $u^\top\Sigma_b u$ and the determinant of the above matrix are non negative. This is met only if $u^\top\Sigma_w u \geq 0$ and $u^\top\Sigma_b u \geq 0$.

\section{Detailed results of the Monte Carlo simulation study}\textbf{}\
In the simulation study the marginal distributions for $X$ and $Y$ are $N(0,1)$,  The copula families  $C_{1,1:n}^{(1)}$ are $C_{1,1:n}^{(3)}$ are normal with respective correlation coefficients $\rho_1 = 0.31$ and $\rho_3 = 0.16$.  This gives Kendall's respectively equal to 0.1 and 0.2 and $\eta$ parameters equal to -0.80 and -1.69.  In the simulations, correlations $\rho$ are parameterized in terms of their logit transform $\eta$. The number of clusters are $m=10$ and $m=50$.  Simulations with equal (EQ) and unequal (NEQ) number of units within each cluster are presented:
\begin{itemize}
    \item For $m=10$, The NEQ scenario has 4 clusters of size $n_i=15$ and 6 of size $n_i=40$ for a total of $\sum n_i=300$ units.  The EQ scenario has $n_i=30$ units in the 10 clusters.
    \item For $m=50$, the NEQ has 30 clusters of size $n_i=10$ and 20 of size $n_i=30$ for a total of $\sum n_i=900$ units thus the EQ scenario has $n_i=18$ units in the 50 clusters.
\end{itemize}

The three bivariate copulas $C_{}^{(2)}$ used in the simulations are the normal copula, the Clayton copula and a Khoudraji copula (20) \citep{genest1998discussion} 
derived from a normal copula. The following tables present the results for the all the parameters of the joint density.  Note that the IFM estimators of $\mu_k$ and $\sigma_k^2$ are, for $k=1,2$, the marginal means and variances of the $X$ and the the $Y$ samples. The maximum likelihood estimators of these parameters have slightly smaller variances, especially when the cluster sample sizes $\{n_i\}$ are unequal. This agrees with the findings for the copulas parameters reported in Section 4.3. 

The R program to obtain the table \ref{table:Resultat de la simulation obtenu scenario2}, \ref{table:Resultat de la simulation obtenu scenario3} and \ref{table:essai3} is subdivided into 2 parts.
\begin{itemize}
    \item The first one gives the program when the copula $C_{}^{(2)}$ is archimedian (normal and Clayton) of a parameter and is in the file named \textit{Simulation-Archimed.R}.
     \item The second one gives the R program when the copula $C_{}^{(2)}$ is a Koudradji copula and is in the file named \textit{Simulation-Koudraji.R}.
\end{itemize}
All the files can be found dans le fichier .zip du nom \textbf{SUBMITT-JVMA-2023.zip}.
 \spacingset{1.5}

\begin{landscape}
\begin{table}[H]
\caption{Expectations of the estimators and their variances multiplied by 10, in parentheses, when $C_{}^{(2)}$ is a normal copula.}
%\centering
%\tiny
\begin{tabular}{lllcccccccc} %\toprule
\toprule
$\tau_2(\eta_2)\,\,\,\,$ & $m$ & Type & Method & $\mu_1=0$ & $\sigma_1=1$ & $\mu_1=0$ &$\sigma_2=1$ & $\eta_1=-0.80$ & $\eta_2$ & $\eta_3=-1.69$\\ 
\toprule
 \multirow{1}{2em}{0.4(0.35)} & \multirow{4}{2em}{10} & \multirow{2}{2em}{NEQ} & MV & 0.00(0.32) &  0.99(0.05)  & 0.01(0.25) & 1.00(0.03)   & -0.90(2.02) & 0.37(0.39) & -1.84(2.54) \\
%&
& & &  IFM & 0.00(0.38)  &  0.99(0.06) & 0.00(0.28) & 0.99(0.04) & -0.96(2.39) & 0.36(0.63) & -1.95(2.71)\\ 
\cline{3-11}
& & \multirow{2}{2em}{EQ} & MV & 0.00(0.32) &  0.99(0.05)  & 0.00(0.25) & 1.00(0.02)    & -0.96(1.90) & 0.35(0.37) & -1.81(2.29) \\
%&
& & &  IFM & 0.00(0.33)  &  0.98(0.05) & 0.00(0.25) & 0.99(0.04) & -1.00(2.20) & 0.34(0.62) & -1.89(2.41)\\ %%%FAIT
 \cline{2-11}
 \cline{2-11}
 & \multirow{4}{2em}{50} & \multirow{2}{2em}{NEQ} & MV & 0.00(0.07) &  1.00(0.01)  & 0.00(0.05) & 1.00(0.01)  & -0.84(0.53) & 0.36(0.11) & -1.75(0.84)\\
%&
& & &  IFM & 0.00(0.09) &  1.00(0.02)  & 0.00(0.07) & 1.00(0.01) & -0.85(0.62) & 0.36(0.16) & -1.78(0.91)\\%%%%%%FAIT%%%%%%%%%%%%%
\cline{3-11}
& & \multirow{2}{2em}{EQ} & MV  & 0.00(0.07) &  1.00(0.01)  & 0.00(0.05) & 1.00(0.01)     & -0.85(0.58) & 0.35(0.12) & -1.74(0.75) \\
  %&
  & & &  IFM & 0.00(0.07) &  0.99(0.01)  & 0.00(0.05) & 1.00(0.01)  & -0.86(0.63) & 0.35(0.16) & -1.76(0.80)\\%%%%%%FAIT%%%%%%%%%%%%%
 \toprule %%%%FAIT
\multirow{1}{2em}{0.6(1.44)}  & \multirow{4}{2em}{10} & \multirow{1}{1em}{NEQ} & MV & 0.01(0.30) &  0.99(0.05)  & 0.00(0.26) & 0.99(0.03)  & -0.90(1.86) & 1.44(0.31) & -1.82(2.33) \\
%&
& & &  IFM & 0.00(0.37)  &  0.99(0.06) & 0.00(0.32) & 0.99(0.05) & -0.97(2.33) & 1.43(0.42) & -1.93(2.50)\\ %%%FAIT----VERIFIER 
\cline{3-11}
& & \multirow{2}{2em}{EQ} & MV & 0.00(0.28) &  1.00(0.05)  & 0.01(0.25) & 0.99(0.03)    & -0.90(1.97) & 1.45(0.31) & -1.83(2.24) \\
%&    
& & &  IFM & 0.00(0.30)  &  0.99(0.06) & 0.01(0.26) & 0.99(0.04) & -0.95(2.30) & 1.44(0.41) & -1.93(2.50) \\ %%%FAIT
\cline{2-11}  
  & \multirow{4}{2em}{50} & \multirow{2}{2em}{NEQ} & MV & 0.00(0.07) &  1.00(0.01)  & 0.00(0.06) & 1.00(0.01)   & -0.84(0.55) & 1.45(0.09) & -1.74(0.76) \\
%&
& & &  IFM & 0.00(0.09) &  1.00(0.02)  & 0.00(0.08) & 1.00(0.01) & -0.85(0.64) & 1.45(0.12) & -1.77(0.84)\\ %%%A VERIFIER
\cline{3-11}
& & \multirow{2}{2em}{EQ} & MV & 0.00(0.06) &  1.00(0.01)  & 0.00(0.05) & 1.00(0.01)   & -0.84(0.53) & 1.44(0.09) & -1.73(0.70) \\
%&
& & &  IFM & 0.00(0.07) &  1.00(0.01)  & 0.00(0.06) & 1.00(0.01) & -0.85(0.60) & 1.44(0.11) & -1.75(0.76)\\
 \toprule
\end{tabular}
\label{table:Resultat de la simulation obtenu scenario2}
\end{table}
\end{landscape}

\begin{landscape}
\begin{table}[H]
\caption{Expectations of the estimators and their variances multiplied by 10 in parentheses when $C_{}^{(2)}$ is Clayton copula.}
\centering
%\tiny
\begin{tabular}{lllcccccccc}
\toprule
$\tau_2(\delta_2)\,\,\,\,$ & $m$ & Type & Method & $\mu_1=0$ & $\sigma_1=1$ & $\mu_1=0$ &$\sigma_2=1$ & $\eta_1=-0.80$ & $\delta_2$ & $\eta_3=-1.69$\\ 
\toprule
 \multirow{1}{2em}{0.4(1.33)} & \multirow{4}{2em}{10} & \multirow{2}{2em}{NEQ} & MV & 0.00(0.29) &  0.99(0.04)  & 0.00(0.19) & 0.99(0.03)    & -0.90(1.59) & 1.34(0.60) & -1.77(2.17) \\
%&
& & &  IFM & 0.00(0.40)  &  0.98(0.06) & 0.00(0.24) & 0.99(0.05) & -0.99(2.19) & 1.33(1.00) & -1.90(2.51)\\ 
\cline{3-11}
& & \multirow{2}{2em}{EQ} & MV & 0.00(0.30) &  0.99(0.04)  & 0.00(0.22) & 1.00(0.03)    & -0.88(1.63) & 1.35(0.66) & -1.82(2.36) \\
%&  
& & &  IFM & -0.01(0.34)  &  0.98(0.06) & 0.00(0.23) & 0.99(0.04) & -0.98(2.18) & 1.33(0.84) & -1.91(2.68)\\ %%%FAIT
 \cline{2-11}
  & \multirow{4}{2em}{50} & \multirow{2}{2em}{NEQ} & MV & 0.00(0.07) &  1.00(0.01)  & 0.00(0.05) & 1.00(0.01)     & -0.84(0.46) & 1.33(0.18) & -1.73(0.78)\\
%&   
& & &  IFM & 0.00(0.10) &  1.00(0.02)  & 0.00(0.07) & 1.00(0.01) & -0.85(0.67) & 1.33(0.26) & -1.76(0.86)\\%%%%%%FAIT%%%%%%%%%%%%%
\cline{3-11}
& & \multirow{2}{2em}{EQ} & MV  & 0.00(0.06) &  1.00(0.01)  & 0.00(0.04) & 1.00(0.01)     & -0.83(0.46) & 1.33(0.15) & -1.73(0.67) \\
  %&  
  & & &  IFM & 0.00(0.07) &  0.99(0.01)  & 0.00(0.05) & 1.00(0.01)  & -0.85(0.60) & 1.32(0.19) & -1.75(0.73)\\%%%%%%FAIT%%%%%%%%%%%%%
 \toprule %%%%FAIT
\multirow{1}{2em}{0.6(3)}  & \multirow{4}{2em}{10} & \multirow{2}{2em}{NEQ} & MV & 0.00(0.23) &  1.00(0.03)  & -0.01(0.19) & 1.00(0.03)  & -0.86(1.57) & 3.05(2.53) & -1.79(2.52) \\
& & &  IFM & -0.00(0.36)  &  0.98(0.05) & -0.01(0.28) & 0.99(0.05) & -0.98(2.18) & 2.98(3.51) & -1.91(2.58)\\ 
\cline{3-11}

& & \multirow{2}{2em}{EQ} & MV & 0.01(0.22) &  1.00(0.03)  & 0.00(0.20) & 1.00(0.03)    & -0.88(1.48) & 3.02(2.39) & -1.80(2.48) \\
& & &  IFM & 0.01(0.33)  &  0.99(0.05) & 0.01(0.26) & 0.99(0.05) & -0.98(2.11) & 2.96(3.13) & -1.91(2.56)\\ %%%FAIT
\cline{2-11}
  & \multirow{4}{2em}{50} & \multirow{2}{2em}{NEQ} & MV & 0.00(0.05) &  1.00(0.01) & 0.00(0.05) & 1.00(0.01)     & -0.84(0.45) & 3.01(0.63) & -1.76(0.83) \\
%&
& & &  IFM & 0.00(0.09) &  1.00(0.02)  & 0.00(0.07) & 1.00(0.01) & -0.86(0.65) & 3.00(0.91) & -1.79(0.89)\\
\cline{3-11}
& & \multirow{2}{2em}{EQ} & MV & 0.00(0.05) &  1.00(0.01)  & 0.00(0.05) & 1.00(0.01)     & -0.82(0.45) & 3.01(0.61) & -1.73(0.71) \\
%&
& & &  IFM & 0.00(0.07) &  1.00(0.01)  & 0.00(0.06) & 1.00(0.01) & -0.86(0.59) & 2.99(0.77) & -1.75(0.77)\\
\toprule
\end{tabular}
\label{table:Resultat de la simulation obtenu scenario3}
\end{table}
\end{landscape}
\begin{landscape}
\begin{table}[H] 
\caption{Expectations of the estimators and their variances multiplied by 10 in parentheses when $C_{}^{(2)}$ is Khoudraji copula.}
\centering \small
\begin{tabular}{lllccccccccc}
\toprule 
$\tau_2\,\,\,\,$ & $m$ & Type & Method & $\mu_1$ & $\sigma_1$ & $\mu_1$ & $\sigma_2$ & $\eta_1=-0.80$ & $\eta_{\rho}=0.75$ & $\eta_{\kappa}=1.52$ & $\eta_3=-1.69$\\ 
\toprule
 \multirow{1}{2em}{0.4} & \multirow{4}{2em}{10} & \multirow{1}{2em}{NEQ } & MV & 0.00(0.32) &  1.00(0.05)  & 0.00(0.21) & 1.00(0.03)    & -0.89(1.71) & 0.81(0.48) & 1.50(3.55) & -1.78(2.54) \\
& & &  IFM &  0.00(0.40)  &  0.99(0.07) & 0.00(0.27) & 1.00(0.04) & -0.96(2.35)& 0.79(0.61) & 1.46(4.61) & -1.90(2.71)\\
\cline{3-12}  %
& & \multirow{2}{2em}{EQ} & 
        MV & 0.01(0.32) &  0.99(0.05)  & 0.00(0.23) & 0.99(0.02)   & -0.90(1.83) & 0.80(0.55) &  1.51(2.42) & -1.85(2.13) \\
& & &  IFM & 0.01(0.33)  &  0.99(0.06) & 0.00(0.24) & 0.99(0.03) & -0.95(2.08)& 0.75(0.65) & 1.49(3.25) &  -1.89(2.49)\\ %%%FAIT
 \cline{2-12}
 
  & \multirow{4}{2em}{50} & \multirow{2}{2em}{NEQ} & MV & 0.00(0.07) &  1.00(0.01)  & 0.00(0.05) & 1.00(0.01)  & -0.84(0.47) & 0.77(0.16) &  1.52(1.29) & -1.76(0.73) \\
  % %& 
  & & &  IFM & 0.00(0.09) &  0.99(0.02)  & 0.00(0.06) & 0.99(0.01)  & -0.87(0.56) & 0.76(0.18) & 1.53(1.94) & -1.80(0.81)\\%%%%%%FAIT%%%%%%%%%%%%%
\cline{3-12}
%%4
& & \multirow{2}{2em}{EQ} & MV  & 0.01(0.07) &  1.00(0.01)  & 0.01(0.05) & 1.00(0.01)   & -0.82(0.51) & 0.77(0.12) & 1.51(0.98) & -1.73(0.63) \\
  %&
  & & &  IFM & 0.01(0.08) &  1.00(0.01)  & 0.01(0.05) & 1.00(0.01)  & -0.83(0.58) & 0.76(0.18) & 1.51(1.44) & -1.74(0.79)\\
 \toprule %%%%FAIT 
 &&&&&&&&&&\\
 % \toprule %%%%FAIT
  $\tau_2\,\,\,\,$ & $m$ & Type  & Method & $\mu_1$ & $\sigma_1$ & $\mu_1$ &$\sigma_2$ & $\eta_1=-0.80$ & $\eta_{\rho}=1.45$ & $\eta_{\kappa}=3.48$ & $\eta_3=-1.69$\\ 
  \toprule %%%%FAIT
 \multirow{1}{2em}{0.6} & \multirow{4}{2em}{10} & \multirow{2}{2em}{NEQ} & MV & 0.02(0.33) &  1.00(0.05)  & 0.01(0.27) & 1.00(0.03) &   -0.88(1.71) & 1.51(0.32) & 3.25(6.31) & -1.84(2.25) \\
% %& 
 & & &  IFM & 0.01(0.40)  &  0.98(0.06) & 0.00(0.35) & 0.98(0.05) & -0.99(2.04) & 1.47(0.44) & 3.18(7.26) &  -1.94(2.59)\\
 \cline{3-12}
 & & \multirow{2}{2em}{EQ} & MV & 0.00(0.32) &  1.00(0.05)  & 0.00(0.23) & 1.00(0.03)  & -0.86(1.77) & 1.49(0.36) & 3.39(5.30)  & -1.81(2.20) \\
% %&
 & & &  IFM & 0.00(0.36)  &  0.99(0.06) & -0.01(0.26) & 0.99(0.04) & -0.94(2.09) & 1.45(0.38) & 3.62(6.34) & -1.88(2.74)\\ %%%FAIT.
\cline{2-12}
   & \multirow{4}{2em}{50} & \multirow{2}{2em}{NEQ} & MV & 0.01(0.05) &  1.00(0.01)  & 0.01(0.05) & 1.00(0.01)  & -0.84(0.51) & 1.46(0.08) & 3.56(3.55) & -1.76(0.75)\\
% %& 
 & & &  IFM & 0.01(0.07) &  1.00(0.02)  & 0.00(0.07) & 1.00(0.01)  & -0.84(0.64) & 1.46(0.11) & 3.57(5.37) &  -1.80(0.83)\\
 \cline{3-12}
% %%8
% %&
 & & \multirow{2}{2em}{EQ} & MV & 0.01(0.06) &  1.00(0.02)  & 0.01(0.04) & 1.00(0.01)  & -0.81(0.49) & 1.47(0.11) & 3.48(2.14) & -1.77(0.73)\\
% %& 
 &&& IFM & 0.01(0.06) &  1.00(0.02)  & 0.01(0.04) & 1.00(0.01) & -0.83(0.63) & 1.47(0.14) & 3.50(2.79) & -1.79(0.86)\\
\toprule % 
\end{tabular}
\label{table:essai3}
\end{table}

\end{landscape}

% \section{R program used to process the data in section 5.}
% The data treated in the section are already treated by \cite{Gold2011} with a linear mixed model. These data are contained in the file named \textbf{data-School.csv}.
% To obtain each of the results in Section 5, we used step-by-step fitting R programs whose files reflect the usage.
% \begin{itemize}
%     \item \textbf{Step 1 of fitting:} the program for fitting the marginal laws of $F_0$ and $G_0$ in the file named \textit{Step1: Fitting-marginal-F0-G0.R}

%     \item \textbf{Step 2 of fitting :} the program of fitting of the copula $C_{1,1:n}^{(1)}$ in the file named \textit{Step2: Fitting-Copula-C1.R}.
    
%     \item \textbf{Step 3 of fitting :} the fitting program of the copula $C^{(2)}$ in the file named \textit{Step3: Fitting-Copula-C2.R}
    
%     \item \textbf{Step 4 of fitting :} the fitting program of the copula $C_{1,1:n}^{(3)}$ in the file named \textit{Step4: Fitting-Copula-C3.R}

%     \item \textbf{Step 5 of global model fitting:} the general maximum likelihood model fitting program in the file named \textit{Step5: Overall-estimate.R}
% \end{itemize}
% All the files can be found dans le fichier .zip du nom \textbf{SUBMITT-JVMA-2023.zip}. A Github link can be proposed afterwards to find the elements and the data.%on the link Github \url{https://github.com/Gabatsarl/Submitt-JVMA.git}.

\bibliographystyle{agsm}

\bibliography{Bibliography-TAG}